\documentclass[11pt, a4paper]{article} 

\usepackage[german, english]{babel}
\usepackage[latin1]{inputenc}
\usepackage[T1]{fontenc}
\usepackage{lmodern}
\usepackage{amsmath}    
\usepackage{amsthm}  
\usepackage{amsfonts}     
\usepackage{amssymb}
\usepackage{array}
\usepackage[affil-it]{authblk}
\usepackage{caption}
\usepackage{color}
\usepackage{courier}
\usepackage{diagbox}
\usepackage{dsfont}			
\usepackage{enumerate}			 
\usepackage{epsfig}
\usepackage{float}      
\usepackage{geometry} 
\usepackage{graphicx}   
\usepackage{cite}
\usepackage{listings}        
\usepackage{lastpage} 
\usepackage{mdframed}
\usepackage{multicol}  
\usepackage{newclude}
\usepackage{placeins}
\usepackage{scalerel}
\usepackage{scrpage2}
\usepackage{setspace}
\usepackage{shuffle}
\usepackage{slashed}
\usepackage{subcaption}
\usepackage[tc]{titlepic}
\usepackage{tikz}
\usepackage{tikz-cd}
\usetikzlibrary{arrows,arrows.meta,shapes,trees,decorations.pathmorphing,decorations.markings,backgrounds,calc,positioning}
\usepackage{titlesec}
\usepackage{url}
\usepackage{varwidth}
\usepackage{verbatim}
\usepackage{wrapfig}
\usepackage{xcolor}
\usepackage{CJK}
\usepackage{hyperref}

\newcommand*{\scalar}{ \includegraphics[height=0.5\baselineskip]{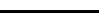} }
\newcommand*{\dashed}{ \includegraphics[height=0.5\baselineskip]{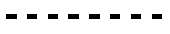} }
\newcommand*{\dotted}{ \includegraphics[height=0.5\baselineskip]{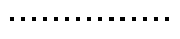} }

\newcommand{\cp}[3][\Gamma]{ \chi_{#1}^{(#2|#3)} }

\newcommand{\kp}[1][\Gamma]{ \Psi_{#1} }

\newcommand{\nquad}{\!\!\!\!}

\newcommand{\Chi}{ \mathrm{X} }

\newcommand*{\defeq}{\mathrel{\vcenter{\baselineskip0.5ex \lineskiplimit0pt
                     \hbox{\scriptsize.}\hbox{\scriptsize.}}}%
                     =}

\DeclareMathOperator{\tr}{tr}

\DeclareMathOperator{\sym}{sym}

\newcolumntype{L}[1]{>{\raggedright\let\newline\\\arraybackslash\hspace{0pt}}m{#1}}
\newcolumntype{C}[1]{>{\centering\let\newline\\\arraybackslash\hspace{0pt}}m{#1}}
\newcolumntype{R}[1]{>{\raggedleft\let\newline\\\arraybackslash\hspace{0pt}}m{#1}}

\makeatletter
\newsavebox{\@brx}
\newcommand{\llangle}[1][]{\savebox{\@brx}{\(\m@th{#1\langle}\)}%
	\mathopen{\copy\@brx\kern-0.5\wd\@brx\usebox{\@brx}}}
\newcommand{\rrangle}[1][]{\savebox{\@brx}{\(\m@th{#1\rangle}\)}%
	\mathclose{\copy\@brx\kern-0.5\wd\@brx\usebox{\@brx}}}
\makeatother

\newcommand{\al}[2][]{\begin{align#1}#2\end{align#1}}

\newcommand{\sa}{\mathsf{a}}

\newcommand{\su}{\mathsf{u}}
\newcommand{\sv}{\mathsf{v}}
\newcommand{\sw}{\mathsf{w}}

\newcommand{\sA}{\mathsf{A}}
\newcommand{\sD}{\mathsf{D}}

\newcommand{\cC}{\mathcal{C}}
\newcommand{\cD}{\mathcal{D}}
\newcommand{\cE}{\mathcal{E}}
\newcommand{\cP}{\mathcal{P}}

\newcommand{\NN}{\mathbb{N}}
\newcommand{\ZZ}{\mathbb{Z}}

\newtheorem{theo}{Theorem}[section]
\newtheorem{exam}[theo]{Example}

\newtheorem{defi}[theo]{Definition}
\newtheorem{prop}[theo]{Proposition}
\newtheorem{cor}[theo]{Corollary}
\newtheorem{rem}[theo]{Remark}
\newtheorem{conj}[theo]{Conjecture}

\newcommand{\reffig}[1]{fig. \ref{#1}}
\newcommand{\refeq}[1]{eq. (\ref{#1})}

\let\svthefootnote\thefootnote
\newcommand\blankfootnote[1]{%
  \let\thefootnote\relax\footnotetext{#1}%
  \let\thefootnote\svthefootnote%
}

\begin{document}

	\title{\textbf{Contraction of Dirac matrices via chord diagrams}}
	\date{}
	\author{Marcel Golz}
	\affil{Institut f\"ur Physik, Humboldt-Universit\"at zu Berlin\\ Newtonstraße 15, D-12489 Berlin, Germany}
	\maketitle
	\thispagestyle{empty}

	\begin{abstract}
		Chord diagrams and combinatorics of word algebras are used to model products of Dirac matrices, their traces, and contractions. A simple formula for the result of arbitrary contractions is derived, simplifying and extending an old contraction algorithm due to Kahane. This formula is then used to express the Schwinger parametric integrand of a QED Feynman integral in a much simplified form, with the entire internal tensor structure eliminated. Possible next steps for further simplification, including a specific conjecture, are discussed.
	\end{abstract}
	
	\tableofcontents

	\blankfootnote{\texttt{mgolz@physik.hu-berlin.de}}

\newpage
\section{Introduction}

\subsection{Motivation}
	The contraction of Dirac matrices is a problem that has in principle been solved since the early 1960s and with the advancement of computers contractions involving any number of matrices can be computed quickly. However, recent work has laid bare the need for an improved understanding of the fundamental combinatorics governing these contractions. In \cite{Golz_2017_CyclePol} we gave an explicit formula for the Schwinger parametric integrand of a Feynman graph $\Gamma$ in quantum electrodynamics. It takes on the form $\gamma_{\Gamma}N_{\Gamma}S_{\Gamma}$, where $\gamma_{\Gamma}$ contains the (traces of) products of Dirac matrices, $S_{\Gamma} = \exp(-\Phi_{\Gamma}/\kp)/\kp^2$ is the usual scalar integrand involving the Kirchhoff and Symanzik polynomials and
	\al{
		N_{\Gamma} \sim \sum_P \left(\prod_{(i,j)\in P} g_{\mu_{i}\mu_{j}}\frac{\cp{i}{j}}{2\kp}\right)\prod_{k\notin P}\frac{\Chi_{\Gamma}^{k,\mu_k}}{\kp}. \label{eq_qed_int}
	}
	Here $\cp{i}{j}$ are the cycle polynomials in Schwinger parameters $\alpha_e$ introduced in \cite{Golz_2017_CyclePol}, $\Chi_{\Gamma}^{k,\mu_k}$ are certain linear combinations of external momenta of $\Gamma$ with cycle polynomials as coefficients and the sum is over all pairings of fermion edges and vertices of the graph\footnote{For example, $P=\{ (1,3) \}$ is a pairing of the set $\{1,2,3,4\}$ with $2,4\notin P$. Note that we omitted some notational technicalities for the sake of clarity, hence  ``$\sim$'' instead of ``$=$''. Details can be found in \cite{Golz_2017_CyclePol}.}. The number of pairings grows factorially with the number of edges and vertices, so an enormous amount of contractions has to be computed. While this is in principle doable via computer algebra, even for large graphs, it is advantageous to understand the contraction combinatorially for a different reason. Knowing the integer coefficients resulting from contraction, it should be possible to exploit the multitude of identities for graph polynomials to express the integrand in a simpler form, with all metric tensors, external momenta and Dirac matrices fully contracted and in section \ref{sec_applic} we give a conjecture for what this should look like specificially. This would make quantum electrodynamics pliable for many mathematical tools previously only applied in scalar theories \cite{Panzer_2014_Algorithms, KompanietsPanzer_2016}.\\

There are a multitude of modern methods that have been developed to deal with the problem of overly complicated contractions (e.g. spin-helicity, BCFW recursion \cite{FengLuo2012, Britto2005, HuangElvang2015}) and the reader may not yet be convinced that studying the combinatorics of the ``traditional'' contraction process is a worthwhile enterprise. However, especially outside of supersymmetric theories, such on-shell methods are not immune to becoming complicated and tedious either, and the standard contraction of Dirac matrices is still very much used today (e.g. in \cite{Gracey_SymmQCD, ChetyrkinZoller_2017}). Instead of circumventing the contraction process, like these methods, we completely work it out, in a way that does not depend on any particular choice of representation for the gamma matrices or spinor basis, and give its end result for any QED graph, at any loop-order, in terms of simple chord diagrams. Moreover, while the direct application of this article's results to scattering amplitude computations is certainly possible, it is hardly its main purpose. Our focus lies much more on the study of Feynman amplitudes (their geometry, number theoretic content etc.) in the parametric context, in which the above methods are plainly not applicable.

\subsection{Dirac matrices, Chisholm's identities and Kahane's algorithm}\label{sec_intro_2}

\begin{figure}[h] \begin{center}
		\includegraphics[scale=1.0]{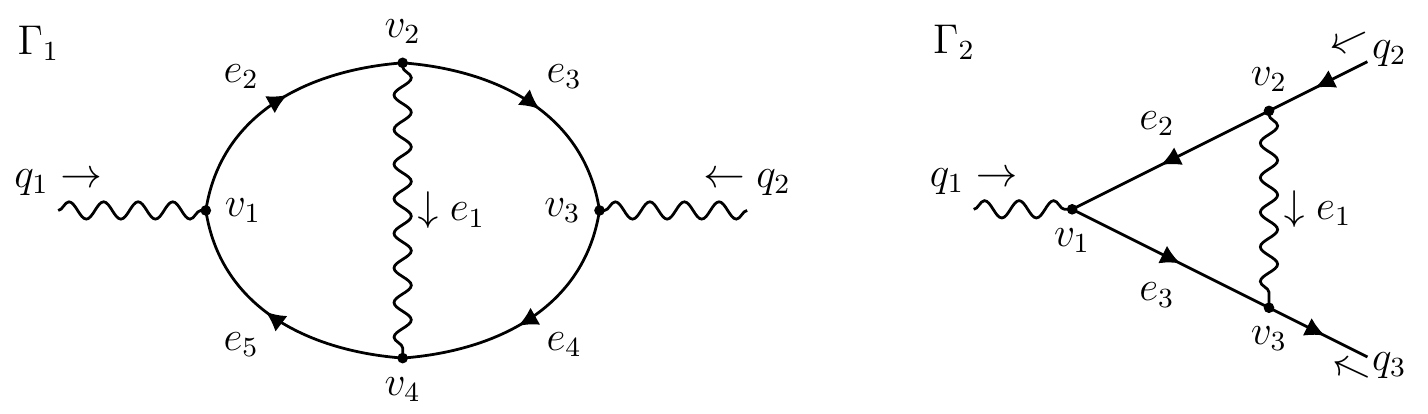}
		\caption[Feynman graphs]{Two examples of Feynman graphs from quantum electrodynamics.}
		\label{01_FeynmanGraphs}
	\end{center}\end{figure}

	The Dirac gamma matrices are a set of four complex $4\times 4$ matrices that satisfy the anticommutation relations
	\al{
		\gamma^{\mu}\gamma^{\nu} + \gamma^{\nu}\gamma^{\mu} = 2g^{\mu\nu}\mathds{1}_{4\times4}\qquad \mu, \nu = 0,1,2,3 \label{eq_Clifford_Gamma}
	}
	and hence generate a representation of a Clifford algebra. In quantum electrodynamics Dirac matrices appear as a consequence of the Feynman rules (which, in turn, are motivated by solutions of the Dirac equation), assigning them to fermion edges and vertices. Overall, one finds for a QED Feynman graph $\Gamma$ that its Dirac matrix structure is a product of an odd number of Dirac matrices, corresponding to the edges and vertices in a path leading from an outgoing to an incoming external fermion edge, and a trace of Dirac matrices for each closed fermion cycle. Consider for example $\Gamma_1$ and $\Gamma_2$ from \reffig{01_FeynmanGraphs}. $\Gamma_1$ only contains a fermion cycle, so one has
	\al{
		\gamma_{\Gamma_1} = \tr( \gamma^{\nu_4}\gamma^{\mu_4}\gamma^{\nu_3}\gamma^{\mu_3}\gamma^{\nu_2}\gamma^{\mu_2}\gamma^{\nu_1}\gamma^{\mu_5}),
	}
	where we use the convention that space time indices $\nu_i$ correspond to vertices $v_i$ and $\mu_i$ to edges $e_i$. For $\Gamma_2$ one only has a fermion path, so
	\al{
		\gamma_{\Gamma_2} = \gamma^{\nu_3}\gamma^{\mu_3}\gamma^{\nu_1}\gamma^{\mu_2}\gamma^{\nu_2},
	}
	where the product has to be ordered by going opposite the fermion flow. The remaining parts of the Feynman rules result in terms containing combinations of the polynomials and external momenta mentioned above as well as metric tensors $g^{\mu\nu}$, resulting in contraction of some or all of the Dirac matrices.\\
	
\paragraph{Contracting Dirac matrices the old-fashioned way.} Traditionally the contraction is computed by iteratively applying the Clifford algebra relation \refeq{eq_Clifford_Gamma}, or rather, an identity that can be derived from it:
	\al{
		\gamma_{\mu} \gamma_{\nu_1}\dotsc\gamma_{\nu_n}\gamma^{\mu}  = \begin{cases} -2 \gamma_{\nu_n}\dotsc\gamma_{\nu_1} &\text{ if } n \text{ odd}\\[3mm]
 2 ( \gamma_{\nu_n}\gamma_{\nu_1}\dotsc\gamma_{\nu_{n-1}} + \gamma_{\nu_{n-1}}\dotsc\gamma_{\nu_1} \gamma_{\nu_n}  ) \quad &\text{ if } n \text{ even}	
 	\end{cases}\label{eq_contraction_Gamma}
	}
	It was first proved (independently and with different methods) by Caianello and Fubini \cite{CaianelloFubini_1952_GammaID} and Chisholm \cite{Chisholm_1952_PhD}. After all duplicate indices within one product of Dirac matrices are contracted one can continue by combining traces with the \textit{Chisholm identity}\footnote{Sometimes the previous \refeq{eq_contraction_Gamma} is also called Chisholm identity, but here we will always use the name to refer to \refeq{eq_Chisholm_Gamma}}\cite{Chisholm_1963_TraceProd}
	\al{
		\gamma_{\mu} \tr( \gamma^{\mu}S) = 2(S+\tilde S)\label{eq_Chisholm_Gamma}
	}
	where $S$ is a product containing an odd number of Dirac matrices and $\tilde S$ is the same product reversed. When that identity cannot be applied anymore the remaining traces are expressed in terms of metric tensors with the recursion formula	
	\al{
		\tr(\gamma^{\mu_1}\dotsc\gamma^{\mu_n}) = \sum_{i=2}^n (-1)^ig^{\mu_1\mu_i}\tr(\gamma^{\mu_2}\dotsc \widehat{\gamma^{\mu_i}}\dotsc \gamma^{\mu_n}).\label{eq_recursion_gamma}
	}
 	
	\begin{rem}\label{rem_decomp_Gamma}
		Note that the even case of the contraction relation can alternatively be expressed in the form
		\al{
			\gamma_{\mu} \gamma_{\nu_1}\dotsc\gamma_{\nu_n}\gamma^{\mu} = 2 ( \gamma_{\nu_{k+1}}\dotsc\gamma_{\nu_n}\gamma_{\nu_1}\dotsc\gamma_{\nu_k} + \gamma_{\nu_k}\dotsc\gamma_{\nu_1} \gamma_{\nu_n}\dotsc\gamma_{\nu_{k+1}}  )
		}
		for any odd $k<n$. This is discussed in more detail in section \ref{sec_diracWords}. To sum up the findings of that section in as condensed a form as possible: As a consequence of the equivalence of different choices of decomposition in the even contraction relation the recursive trace formula \refeq{eq_recursion_gamma} reduces to a much shorter, non-recursive formula from which - among other things - the Chisholm identity \refeq{eq_Chisholm_Gamma} follows as a trivial special case. This simplification in turn allows for the combinatorial interpretation of contraction in section \ref{sec_diagramContraction}.
	\end{rem}
	
	\begin{exam}\label{EX_oldContr}
		Consider contraction of $\gamma_{\Gamma_1}$ with two metric tensors:
		\al{\nonumber
			g_{\nu_2\nu_4}g_{\mu_2\mu_4} \gamma_{\Gamma_1}&= \tr( \underbrace{\gamma^{\nu_2}\gamma^{\mu_2}\gamma^{\nu_3}\gamma^{\mu_3}\gamma^{\nu_2}}_{=-2\gamma^{\mu_3}\gamma^{\nu_3}\gamma^{\mu_2}}\gamma^{\mu_2} \gamma^{\nu_1}\gamma^{\mu_5})\\\nonumber
			&= -2\tr(\gamma^{\mu_3}\gamma^{\nu_3}\underbrace{\gamma^{\mu_2}\gamma^{\mu_2}}_{=4} \gamma^{\nu_1}\gamma^{\mu_5})\\
			&= -32 (g_{\mu_3\nu_3}g_{\nu_1\mu_5} - g_{\mu_3\nu_1}g_{\nu_3\mu_5} + g_{\mu_3\mu_5}g_{\nu_1\nu_3})
		}
		Traces can be combined as follows:
		\al{\nonumber
			\tr(\gamma^{\mu_1}\gamma^{\mu_2}\gamma^{\nu_1}\gamma^{\nu_2})\tr(\gamma^{\mu_1}\gamma^{\mu_2}\gamma^{\nu_3}\gamma^{\nu_4}) &= \tr( \underbrace{\gamma^{\mu_1} \tr(\gamma^{\mu_1}\gamma^{\mu_2}\gamma^{\nu_1}\gamma^{\nu_2})}_{=2(\gamma^{\mu_2}\gamma^{\nu_1}\gamma^{\nu_2}+\gamma^{\nu_2}\gamma^{\nu_1}\gamma^{\mu_2})} \gamma^{\mu_2}\gamma^{\nu_3}\gamma^{\nu_4})\\\nonumber
			&= 2\big( \tr(\underbrace{\gamma^{\mu_2}\gamma^{\nu_1}\gamma^{\nu_2}\gamma^{\mu_2}}_{\substack{=2(\gamma^{\nu_2}\gamma^{\nu_1}+\gamma^{\nu_1}\gamma^{\nu_2})\\ =4g^{\nu_1\nu_2}\hspace{16mm}}}\gamma^{\nu_3}\gamma^{\nu_4}) + \tr(\gamma^{\nu_2}\gamma^{\nu_1}\underbrace{\gamma^{\mu_2}\gamma^{\mu_2}}_{=4}\gamma^{\nu_3}\gamma^{\nu_4})\big)\\\nonumber
			&= 8 \big( g^{\nu_1\nu_2}\tr(\gamma^{\nu_3}\gamma^{\nu_4})  + \tr(\gamma^{\nu_2}\gamma^{\nu_1} \gamma^{\nu_3}\gamma^{\nu_4})\big)\\
			&= 32\big( 2g^{\nu_1\nu_2}g^{\nu_3\nu_4} - g^{\nu_1\nu_4}g^{\nu_2\nu_3} + g^{\nu_1\nu_3}g^{\nu_2\nu_4}\big)
		}
	\end{exam}
	
\paragraph{Algorithmic contraction.} 
	Computer algorithms for contraction (e.g. implemented as \texttt{trace4} in \texttt{FORM} \cite{Vermaseren_FormNew}) typically try to successively apply the three equations (\ref{eq_contraction_Gamma}), (\ref{eq_Chisholm_Gamma}) and (\ref{eq_recursion_gamma}) until full contraction is achieved. However, as far back as the 1960s there have been attempts to find alternative contraction methods that bear some similarities to our approach \cite{Kahane_1967_GammaAlgorithm}. Kahane developed an algorithm which involves instructions on how to first draw a diagram based on a given sequence of Dirac matrices. Following that the algorithm describes how to parse the diagram, simultaneously multiplying the result with certain factors depending on what one encounters. In our approach we use chord diagrams - a very well understood type of graph - together with a colouring to carry all the necessary information. Moreover, we isolate the relevant combinatorial property of the chord diagrams - the number of cycle subgraphs with a certain colouring - such that our result is a closed formula instead of an algorithm. Finally, Kahane's proofs are based on using a certain basis for the Clifford algebra generated by the Dirac matrices, while our results are entirely concluded from the contraction relation \refeq{eq_contraction_Gamma}. In fact, in section \ref{sec_diracWords} we completely abstract the process of contraction from Dirac matrices to combinatorial sequences of letters representing the different space-time indices.\\

Kahane's algorithm was later generalised to products of traces by Chisholm \cite{Chisholm_1972_GammaAlgGen}, using his identity \refeq{eq_Chisholm_Gamma}. Working with Kahane diagrams the computations with this generalised algorithm become quite cumbersome\footnote{In the words of J.S.R. Chisholm himself \cite{Chisholm_1972_GammaAlgGen}
: \textit{``The proof of our final result is long and tedious, and even the statement of it is fraught with notational difficulties. We therefore explain it by an example, [...]''}}. Following our approach the general case follows very directly and with only marginally more complicated notation as corollary \ref{cor_generalisation} from our single trace result theorem \ref{theo_contraction_diagrams}.

\section{From Dirac matrices to words}\label{sec_diracWords}

\subsection{The algebra of Dirac words}\label{sec_diracWords_intro}
	In this section we define an algebra that will serve as an abstraction of products of Dirac matrices and allow us to study their contraction and traces without any of the unnecessary ballast they carry.\\
	
	Let $\sA \defeq \{ \sa_i\ | \ i \in \NN \}$ be an alphabet. Then $\sA^*$ with $^*$ denoting the Kleene star \cite{Kleene_1967_Logic} is the set of words $\sw$ (``noncommutative monomials'') over $\sA$. The length, i.e. the number of letters, of a word $\sw$ is denoted $|\sw|$. We say a word is even (odd) if its length is even (odd) and we only consider words of finite length. $\tilde{\sw}$ is the reversed word. Evidently, $\sA^*$ is a free monoid. Moreover, $\sA$ generates a free algebra $\ZZ\langle \sA\rangle$ and we also use the nomenclature ``word'' for elements $\sw = \sum c_j \sw_j$ of this algebra. Unless explicitly stated otherwise we consider homogeneous words in which all ``monomial words'' have the same coefficient and are just rearrangements of the same letters. By linearity the discussion below holds in general, but we will see that we are only really interested in this kind of word.\\
	
	In order to model Dirac matrices we have to satisfy three additional conditions:
	\begin{itemize}
		\item Each space-time index (i.e. each letter $\sa_i\in \sA$) appears at most twice.
		\item An analogon of the contraction relation \refeq{eq_contraction_Gamma} holds.
		\item The word $\delta_{ij}  \defeq \frac{1}{2}(\sa_i\sa_j + \sa_j\sa_i)\in \ZZ\langle \sA\rangle$ has the right properties to serve as an analogon for the metric tensor.
	\end{itemize}
	
	We implement the first condition in our definition of Dirac words.
	\begin{defi}\textbf{\textup{(Dirac words)}}\\
		Let $\sA$ be the alphabet introduced above and $I_k \defeq \langle \sa_i^k \ | \ i\in \NN\rangle$ the ideal generated by $k$-th powers of its letters. Then we define \textup{Dirac words} as elements of the free algebra $\ZZ\langle \sA\rangle$ divided by all third powers
		\al{
			\sD \defeq \ZZ\langle \sA\rangle / I_3.
		}
		Moreover, we define \textup{fully contracted Dirac words} as those Dirac words in which each letter appears at most once, i.e.
		\al{
			\bar \sD \defeq \ZZ\langle \sA\rangle / I_2.
		}
	\end{defi}
	
	\noindent The contraction relation \refeq{eq_contraction_Gamma} is translated to letters and words in the obvious way as
	 \al{
		\sa_i \su \sa_i = -2 \tilde\su \hspace{3cm}
		\sa_i \sa_j\su \sa_i =	 2(\su\sa_j + \sa_j\tilde\su) \label{eq_contraction_words}
	}
	for any odd $\su\in\sD$. In remark \ref{rem_decomp_Gamma} we discussed that the even case can be expressed in different but equivalent ways. We extend this discussion in section \ref{sec_comm_traces} which will allow us to formulate the contraction relation more elegantly in \refeq{eq_contraction_words_elegant}, but for now this version suffices. Note that the even case also includes length 0, i.e. the empty word, as $\sa_i^2 = 2(1+1)=4$. Hence, each letter is up to an integer factor its own multiplicative inverse. This generalises to (monomial) words as $\sw^{-1} = 2^{-2|\sw|}\tilde\sw$. \\

	Finally, we can also introduce an analogue to the metric tensor by simply defining it as an abbreviation for a certain element of $\sD$ that turns out to have exactly the desired properties.
	\begin{prop}
		Let $\delta_{ij} = \frac{1}{2}(\sa_i\sa_j + \sa_j\sa_i) \in \sD$. Then:
		 \al[*]{
			(i)\quad \delta_{ii} = 4 \hspace{2cm} 
			(ii)\quad \delta_{ij} \sa_j = \sa_i\hspace{2cm}
			(iii)\quad \delta_{ij}\sw = \sw\delta_{ij} \ \forall \sw \in \sD
		}
	\begin{proof}
		The first equation follows directly from $\sa_i^2 = 4$. For $(ii)$ we employ the contraction relations (\ref{eq_contraction_words}) to find
		\al{
			\delta_{ij} \sa_j = \frac{1}{2}(\sa_i \sa_j\sa_j + \sa_j\sa_i\sa_j) = \frac{1}{2}(4\sa_i -2\sa_i) = \sa_i.
		}
		In order to prove (iii) note first that the exchange of a letter that we just proved also works if there is a word between $\delta_{ij}$ and $\sa_j$, i.e. for $\su\in\sD$ with $\sa_j\notin \su$
		\al{
			\delta_{ij} \su \sa_j = \frac{1}{2}(\sa_i\sa_j\su\sa_j + \sa_j\sa_i\su\sa_j) = -\sa_i\tilde\su + (\su\sa_i + \sa_i\tilde\su) = \su\sa_i
		}
		if $|\su|$ odd, and 
		\al{
			\delta_{ij} \su \sa_j = \frac{1}{4}\sa_k\sa_i\sa_j\sa_k\su\sa_j = -\frac{1}{2}\sa_k\sa_i\tilde\su\sa_k = \su\sa_i,
		}
		if $|\su|$ even. In the latter case we used \refeq{eq_contraction_words} to rewrite $\delta_{ij}$ as
		\al{
			\delta_{ij} = \frac{1}{2}(\sa_i\sa_j + \sa_j\sa_i) = \frac{1}{4}\sa_k\sa_i\sa_j\sa_k
		}
		for some $k\neq i,j$. This is now used to show commutativity with a single letter, which suffices since a word can be commuted by sequentially commuting its letters:
		\al{
			\delta_{ij}\sa_l& = \frac{1}{4}\sa_k\sa_i\sa_j\sa_k\sa_l
						= -\frac{1}{4}\sa_k\sa_i\sa_j\sa_l\sa_k + \frac{1}{2}\sa_k\sa_i\sa_j\delta_{kl}
						= \frac{1}{2}\sa_l\sa_j\sa_i + \frac{1}{2}\sa_l\sa_i\sa_j
						= \sa_l\delta_{ij}
		}	
	\end{proof}
	\end{prop}
	
	\begin{rem}
	The reader might be wondering why we did not simply use $\delta_{ij} = \frac{1}{2}(\sa_i\sa_j + \sa_j\sa_i)$ as a Clifford algebra equation and derive the contraction relations from there as one does with Dirac matrices. However, that is not possible in this setting. In the Dirac matrix setting one can only derive \refeq{eq_contraction_Gamma} from \refeq{eq_Clifford_Gamma} with the help of the additional information that there are only four Dirac matrices $\gamma_0, \gamma_1, \gamma_2, \gamma_3$ or equivalently the fact that there are four space-time dimensions. Therefore we include \refeq{eq_contraction_words} by definition and derive everything we need from there.
	\end{rem}

\subsection{Commutativity, symmetry equivalence and traces}\label{sec_comm_traces}
\paragraph{Symmetry and equivalence.}

	In this section we discuss some properties of symmetric Dirac words to highlight their importance. Define a symmetrisation/antisymmetrisation map $\sym : \sD \to \sD$ with
	\al{
		\sym(\sw) = \frac{1}{2}(\sw + (-1)^{|\sw|}\tilde\sw).
	}
	such that $\sym(\sD) \subset \sD$ is the subset of even symmetric and odd antisymmetric Dirac words. Let furthermore $s_k: \sD\to\sD$ be the $k$-fold cyclic shift, i.e. for a (monomial) word $\sa_i\sa_j\sv$ one has $s_1(\sa_i\sa_j\sv) = \sa_j\sv\sa_i$, $s_2(\sa_i\sa_j\sv) = \sv\sa_i\sa_j$ and so on. Using this new notation, reconsider the contraction relation \refeq{eq_contraction_words}. The even case is
	 \al{
		\sa_i \sa_j\su \sa_i =	 2(\su\sa_j + \sa_j\tilde\su)  = 4\sym( s_1(\sa_j\su) ).
	}
	We mentioned above that different decompositions are possible. Using the odd case of the contraction relation we find for an even word $\sw = \sv\su$ with $|\sv|, |\su|$ odd that
	\al{\nonumber
		\sa_i \sv\su \sa_i &= -\frac{1}{2} \sa_i \sv\sa_k\tilde\su\sa_k\sa_i\\\nonumber
					&= +\frac{1}{2} \sa_i \sv\sa_k\tilde\su\sa_i\sa_k - \delta_{ik}\sa_i \sv\sa_k\tilde\su\\
					&= -\su\sa_k\tilde\sv\sa_k - \sa_k \sv\sa_k\tilde\su
					= 2(\su\sv + \tilde\sv\tilde\su)
					= 4\sym( s_{|\sv|}(\sw) )
	}
	We see that -- as far as the symmetrisation map is concerned -- all the odd cyclic shifts of even words are the same. In other words:
	\begin{prop}\label{prop_cyclicshift_words}
		Let $\su \in \sD$ be a Dirac word with $|\su|$ even. Then
		\al{
			\sym(\su) = \sym(s_{2k}(\su)) \qquad \forall k\in \NN.  \label{eq_equivalence_symwords}
		}
	\end{prop}
	\noindent The symmetrisation map induces an equivalence relation on $\sD$ given by $\su \sim_{\sym} \sv$ if and only if $\sym(\su) = \sym(\sv)$. For a given even word $\sw$ there are two equivalence classes related by odd cyclic shifts: $[\sw] \defeq \{ s_{2k}(\sw) \ | \ k \in \NN\}$ and $[\sw^*] \defeq \{ s_{2k+1}(\sw) \ | \ k \in \NN\}$. Whenever no confusion can arise, we simply write $\sw, \sw^*$ for (an arbitrary representative of) the equivalence classes, such that odd cyclic shifts become maps $s_{2k+1}(\sw) = \sw^*$ and vice versa. The contraction relation in this notation becomes
	 \al{
		\sa_i \su \sa_i &= \begin{cases} \qquad -2 \tilde\su &\text{ if $|\su|$ odd} ,\\[2mm]
								4\sym(\su^*) &\text{ if $|\su|$ even}. \label{eq_contraction_words_elegant}
					\end{cases}
	}
\paragraph{Commutativity.}
	Above we observed that $\delta_{ij} = \frac{1}{2}(\sa_i\sa_j + \sa_j\sa_i) = \sym(\sa_i\sa_j)$ commutes with all other words. We can generalise this commutation property to longer words as follows.	
	\begin{prop}\label{prop_commute_words}
	Let $\sv,\sw \in \sD$ with $|\sv|$ even. Then
		\al{
			\sw\sym(\sv) = \begin{cases} \ \sym(\sv)\sw &\text{ if } |\sw| \text{ even},\\
				 					\sym(\sv^*)\sw &\text{ if } |\sw| \text{ odd},\end{cases} \label{id_commute_words}
		}
	for all $\sw\in \sD$. Moreover, a word $\su\in \sD$ is a commutative element, i.e. $\su\sw = \sw\su$ for all $\sw \in \sD$, if and only if there exists an even $\sv\in\sD$ such that
		\al{
			\su = \sym(\sv + \sv^*).
		}
	\begin{proof}
		Consider commutation of a letter,
		\al{\nonumber
			\sa_i \sym(\sv) 	= \frac{1}{4}\sa_i \sa_j \sv^*\sa_j 
								= \frac{1}{4}(-\sa_j \sa_i + 2\delta_{ij})\sv^*\sa_j
								= \sym(\sv^*)\mathsf a_i.
		}
		Hence, successively commuting an odd or even number of letters in a word produces the first claim \refeq{id_commute_words} and commutativity of any $\su = \sym(\sv + \sv^*) = \sym(\sv) + \sym(\sv^*)$ is an immediate consequence. To see that all commutative elements have to be of this form consider the following two conditions. If $\su$ is commutative and even then on the one hand
		\al{
			\sa_i \su = \su \sa_i = -\frac{1}{2}\sa_j\sa_i\tilde\su\sa_j = -\frac{1}{2}\sa_j\sa_i\sa_j\tilde\su = \sa_i \tilde\su,
		}
		i.e. $\su \stackrel{!}{=} \tilde\su$. On the other hand one also has 
		\al{
			\sa_i \sym(\su) = \sym(\su) \sa_i = \sa_i\sym(\su^*)
		}
		by commutativity and \refeq{id_commute_words}, so $\su \stackrel{!}{=} \su^*$. Finally, there can be no odd commutative word since that would directly contradict the odd case of \refeq{id_commute_words}.
	\end{proof}
	\end{prop}
	\paragraph{Traces of Dirac words.}
	 We have seen in the beginning that after contraction of all duplicate indices the trace of a product of Dirac matrices is computed with a recursion formula that decomposes it into metric tensors. We can translate that formula to our algebra to define the trace of Dirac words as a linear automorphism
	\al{
		\tr :\ \sa_{i_1}\dotsc\sa_{i_n} \mapsto \sum_{j=2}^n (-1)^j\delta_{i_1i_j}\tr(\sa_{i_2}\dotsc \hat\sa_{i_j}\dotsc \sa_{i_n}), \qquad \forall n\geq 2
	}
	on $\sD$, with the trace of the empty word $\tr(1)\defeq 4$ corresponding to the trace of the $4\times4$ unit matrix in the Dirac matrix case. 
	The trace $\tr(\sw) \in \sD$ is clearly commutative for every $\sw \in \sD$, so by proposition \ref{prop_commute_words} there exists a word $\sw'\in\sD$ such that
	\al{
		\tr(\sw) = \sym(\sw' + \sw'^*)
	}
	and $\sw'$ differs from $\sw$ at most by a constant factor, which we discuss in the following
	\begin{theo}\label{theo_wordtrace}
	For all $\sw \in \sD$ with $|\sw|$ even
	\al{
		\tr(\sw) = 2\sym(\sw + \sw^*).
	}
	\begin{proof}
		For $|\sw| \in \{0,2\}$ we can check explicitly that the claim holds:
		\al{
			2\sym(1+1) = 4 = \tr(1)\\[2mm]
			2(\sym(\sa_i\sa_j) + \sym(\sa_j\sa_i)) = 4\delta_{ij} &= \tr(\sa_i\sa_j)
		}
		Exploiting the recursive trace formula we then show the general case. Consider the word $\sa_1\sa_2\dotsm\sa_n$ and commute the first letter all the way to the end,
		\al{\nonumber
			\sa_1\sa_2\dotsm\sa_n &= - \sa_2\sa_1\sa_3\dotsm \sa_n + 2\delta_{12}\sa_3\dotsm \sa_n\\\nonumber
							& \ \,  \vdots \\
							& = (-1)^{n-1}\underbrace{\sa_2\dotsm\sa_n\sa_1}_{=s_1(\sa_1\dotsm\sa_n)}
								+ 2\sum_{i=2}^n (-1)^i \delta_{1i}\sa_2\dotsm\hat\sa_i\dotsm\sa_n.
		}
		Using $\sw_{ij}$ for the word $\sw$ after removal of the $i$-th and $j$-th letter we can therefore write
		\al{
			2\sum_{i=2}^{|\sw|} (-1)^i \delta_{1i}\sw_{1i} = \sw +(-1)^{|\sw|}s_1(\sw).
		}
		which is $\sw + \sw^*$ for even words. When trying to do the same for a sum $\sum (-1)^i \delta_{1i}(\sw_{1i})^*$ one encounters problems since $(\sw_{ij})^* \neq (\sw^*)_{ij}$. However, exploiting the symmetrisation and proposition \ref{prop_cyclicshift_words} one quickly shows that for an even word $\sw$
		\al{
			2\sum_{i=2}^{|\sw|} (-1)^i \delta_{1i} \sym((\sw_{1i})^*) = \sym(\sw + \sw^*) = 2\sum_{i=2}^{|\sw|} (-1)^i \delta_{1i} \sym(\sw_{1i}).
		}
		The trick is to move each $\delta_{1i}=\frac{1}{2}(\sa_1\sa_i+\sa_i\sa_1)$ into the $i$-th slot of $\sw_{1i}$, i.e. the place where the $i$-the letter has been removed. In the sum on the rhs this leads to a telescopic sum in which only half of the first and last terms remain. Due to the symmetrisation and proposition \ref{prop_cyclicshift_words} the same trick can be applied to the sum with $(\sw_{1i})^*$ albeit with slightly less obvious cancellations. Hence, one recursively finds
		\al{
			\tr(\sw) = \sum_{i=2}^{|\sw|} (-1)^i\delta_{1i}\tr(\sw_{1i})
				= 2\sum_{i=2}^{|\sw|} (-1)^i\delta_{1i} \sym( \sw_{1i} + (\sw_{1i})^* )
				= 2\sym( \sw + \sw^* ).
		}
	\end{proof}
	\end{theo}
	
	\begin{rem}
		With the above expression for traces one immediately sees Chisholm's identity \refeq{eq_Chisholm_Gamma} as a special case:
		\al{
			\sa_i\tr(\sa_i\sw) = 2\sa_i\sym(\sa_i\sw+\sw\sa_i) = \sa_i^2\sw + \sa_i\tilde\sw\sa_i + \sa_i\sw\sa_i + \sa_i^2\tilde\sw
						= 2(\sw+\tilde\sw)
		}
	\end{rem}
	
	\begin{exam}
		Consider the trace of a word of length $6$ which gives 15 terms in its usual expansion:
		\al{\nonumber
			\frac{1}{4}\tr(\sa_1\sa_2\sa_3\sa_4\sa_5\sa_6) 
			&= \delta_{12}\delta_{34}\delta_{56} - \delta_{12}\delta_{35}\delta_{46} + \delta_{12}\delta_{36}\delta_{45}
			 -\delta_{13}\delta_{24}\delta_{56} + \delta_{13}\delta_{25}\delta_{46} - \delta_{13}\delta_{26}\delta_{45}\\\nonumber
			& +\delta_{14}\delta_{23}\delta_{56} - \delta_{14}\delta_{25}\delta_{36} + \delta_{14}\delta_{26}\delta_{35}
			 -\delta_{15}\delta_{23}\delta_{46} + \delta_{15}\delta_{24}\delta_{36} - \delta_{15}\delta_{26}\delta_{34}\\
			& +\delta_{16}\delta_{23}\delta_{45} - \delta_{16}\delta_{24}\delta_{35} + \delta_{16}\delta_{25}\delta_{34}
		}
		On the contrary, our new expression for the trace has only four terms:
		\al{\nonumber
			\tr(\sa_1\sa_2\sa_3\sa_4\sa_5\sa_6) &= 2\sym(\sa_1\sa_2\sa_3\sa_4\sa_5\sa_6 + \sa_2\sa_3\sa_4\sa_5\sa_6\sa_1)\\
			&= \sa_1\sa_2\sa_3\sa_4\sa_5\sa_6 + \sa_2\sa_3\sa_4\sa_5\sa_6\sa_1
				+ \sa_6\sa_5\sa_4\sa_3\sa_2\sa_1 + \sa_1\sa_6\sa_5\sa_4\sa_3\sa_2
		}
		Moreover, this version of the trace has four terms for any length of word, while the number of terms in the recursive expansion grows factorially as $(|\sw|-1)!!$.
	\end{exam}

	We have now completely abstracted the process of computing traces of Dirac matrices to computations on words. However, explicitly applying the contraction relation \refeq{eq_contraction_words_elegant} to reduce a word from $\sD$ to $\bar \sD$ is still tedious and not very insightful from a theoretical viewpoint. In section \ref{sec_diagramContraction} we use the results of this chapter to abstract further to a purely diagrammatical approach. First however we would like to offer a different perspective on Dirac words that may prove useful in future work.
		
 \subsection{A different perspective - Dirac words as Cartier-Foata monoids}\label{sec_diffpersp}
 	Here we give an alternative interpretation of the previous section's content in terms of slightly different combinatorial objects. While this overcomplicates matters for the purposes of this article it offers both surprising connections to other disciplines and potential future application of this article's results. Both the seminal articles \cite{CartierFoata_1969_CommRearr} and \cite{Mazur_1977_Concurr} as well as the books  
\cite{Lothaire_1997_CombWords, Diekert_1990_Traces} are useful resources for more detail.\\

	The idea is to use an alphabet together with so-called \textit{dependency relations} on it to generate a free partially commutative monoid. They were first used by Cartier and Foata in combinatorics \cite{CartierFoata_1969_CommRearr} and later applied in computer science by Mazurkiewicz \cite{Mazur_1977_Concurr}. Following the shorter nomenclature of the latter these objects are often called trace monoids and their elements traces, but in order to avoid confusion with the -- as far as we can tell -- completely unrelated notion of trace that we discuss in this article we will continue to use the longer more explicit name.
	\begin{defi}\textbf{\textup{(Free partially commutative monoid)}}
		Let $\Sigma$ be a finite alphabet and $R\subseteq \Sigma\times\Sigma$ a reflexive and symmetric relation thereon. $R$ generates a congruence $\sim_R$ on $\Sigma^*$. The \textit{free partially commutative monoid on} $\Sigma$ \textit{relative to} $R$ is defined as the quotient monoid
		\al{
			M(\Sigma, R) \defeq \Sigma^*/\sim_R
		}
	\end{defi}
	\noindent What that means explicitly is that for a pair of letters $(\sa, \mathsf b)\in R$ one has the equality $\sa\mathsf b = \mathsf b\sa$. Similarly such a relation defines a free partially commutative algebra $\ZZ\langle \Sigma, R\rangle \defeq \ZZ\langle \Sigma\rangle/I_R$ where one divides the free $\ZZ$-algebra generated by the alphabet by the ideal $I_R = \langle \sa\mathsf b - \mathsf b \sa \ | \ (\sa,\mathsf b) \in R \rangle$ generated by the relation. Alternatively one can also see the same structure as a free partially commutative Lie algebra by interpreting $R$ as generating a Lie ideal $\langle [\sa, \mathsf b] \ | \ (\sa,\mathsf b) \in R \rangle$ and dividing by that \cite{DuchampKrob_1993_FreePartComm}. With just a little bit more effort one can also find two dual Hopf algebra structures on such a free partially commutative algebra \cite{Schmitt_1990_HopfMonoid}.\\

In order to apply this to Dirac words one could now use an alphabet $\Sigma = \{ \sa_i\ | \ i \in \NN \} \cup \{ \delta_{ij}\ | \ i,j\in \NN \}$ in which the $\delta_{ij}$ are not abbreviations for algebra elements but separate letters. Their commutativity is then introduced as a dependency relation. The contraction relations (\ref{eq_contraction_words}) together with $\delta_{ij} = \frac{1}{2}(\sa_i\sa_j+\sa_j\sa_i)$ then generate a confluent and noetherian rewriting system, which in this case is not surprising because that essentially only means the contraction from $\sD$ to $\bar \sD$ can be automated via computer. Such a rewriting system is a generalisation of what would be called a semi-Thue system\footnote{These systems were of enormous importance in the development of formal languages and mathematical logic. The article \cite{Post_1947_thue} for example contains the first ever proof of undecidability of a classical mathematical problem. Semi-Thue systems are also known as monoid presentations (not to be confused with representations) or string-rewriting systems and are isomorphic to both unrestricted grammars and Turing machines \cite{Davis_1994_CCL}.}\cite{Thue_1914_words, Post_1947_thue} in the case of (not partially commutative) free monoids. This alternative interpretation might prove interesting in the future for two reasons. The unexpected connections to computer science by way of combinatorics hint at a vast untapped potential of interdisciplinary collaboration. So far little research has gone into this direction but articles like \cite{Manin_2009_RenCS_I, Manin_2009_RenCS_II}, or \cite{Marcolli_2015_GraphGrammar} -- where it was shown that Feynman graphs can be interpreted as a type of formal language generated by a theory dependent graph grammar -- seem to suggest that there are deep connections between the two fields whose study might benefit both disciplines.\\

Moreover, Cartier and Foata originally introduced their monoids to prove a (noncommutative generalisation of) MacMahon's Master Theorem, which in its simplest form is stated as follows: Let $A=(a_{ij})_{1\leq i,j\leq n}$ be a matrix with entries in a commutative ring and $x_1,\dotsc,x_n$ formal variables. Let furthermore $C(k_1,\dotsc,k_n)$ denote the coefficient of $x_1^{k_1}\dotsm x_n^{k_n}$ in the product $\prod_{i=1}^n( a_{i1}x_1 + \dotsc + a_{in}x_n)^{k_i}$ and $X=(\delta_{ij}x_i)_{1\leq i,j \leq n}$ the diagonal matrix with the formal variables as entries. Then
	\al{
		\sum_{(k_1,\dotsc,k_n)} C(k_1,\dotsc,k_n)x_1^{k_1}\dotsm x_n^{k_n} = \frac{1}{\det(\mathds 1_{n\times n} - XA)}
	}
	where the rhs is to be understood as a formal expansion with 
	\al{
		\det(\mathds 1_{n\times n} - XA)^{-1} = (1-R)^{-1} = 1+ R + R^2 + \dotsc
	} 
	and the sum is over all tuples of non-negative integers. Since the Kirchhoff polynomial and the various other graph polynomials that typically appear in parametric Feynman integrals can all be expressed as determinants of certain matrices, inserting the right matrix for $A$ will yield graph polynomials in the coefficients. This connection between seemingly disparate objects - graph polynomials and free partially commutative monoids - may prove useful when trying to unveil deeper combinatoric structures in quantum field theories.
	
\section{Diagrammatic contraction}\label{sec_diagramContraction}

\subsection{Chord diagrams}\label{sec_chordDiagrams}
	
	A graph $G$ is an ordered pair $(V_G, E_G)$ of the set of vertices $V_G = \{ v_1, \dotsc, v_{|V_G|}\}$ and the set of edges $E_G = \{ e_1, \dotsc, e_{|E_G|}\}$, together with a map $\partial: E_G \to V_G \times V_G$. A cycle is a 2-regular graph, and here we always take cycle to mean \textit{simple} cycle, i.e. having only one connected component.
	\begin{defi}\textup{\textbf{(Chord diagram)}}\label{def_chord_diagram}\\
		A \textup{chord diagram} $D$ of order $n$ is a graph, consisting of a cycle on $2n$ vertices (the \textup{base}) and $k\leq n$ more edges that pairwise connect $2k$ of the vertices of that cycle (the \textup{chords}). We denote with $\cD^n_k$ the set of all chord diagrams of order $n$ with $k$ chords.
	\end{defi}
	
	\begin{figure}[h]
		\begin{center}
			\includegraphics[width=0.95\textwidth]{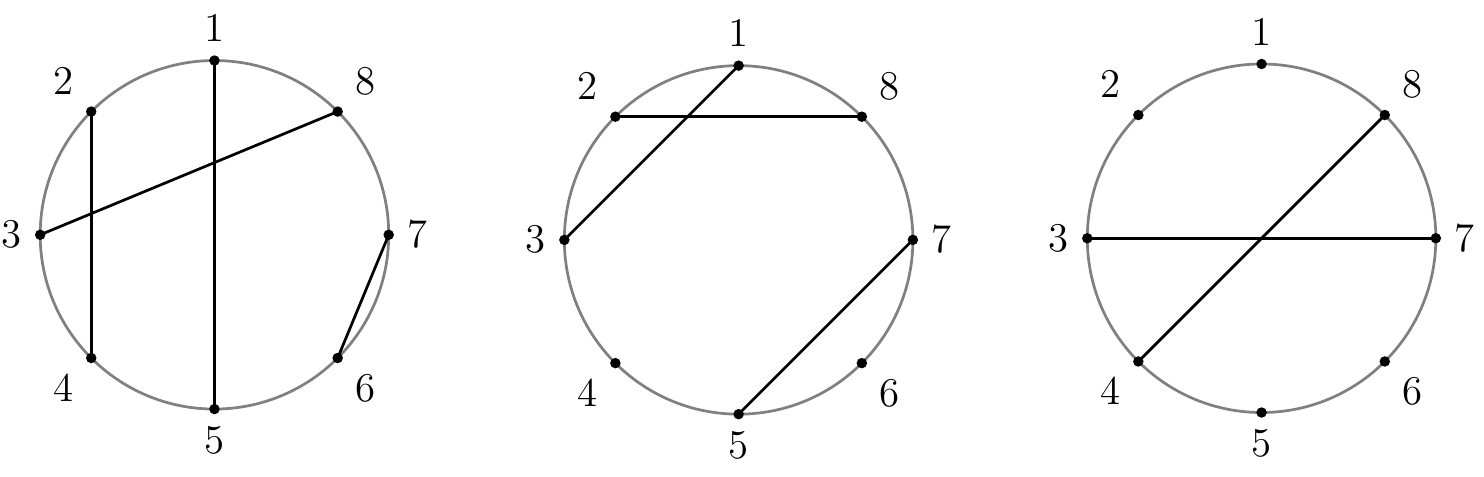}
			\caption{Three chord diagrams of order $n=4$ with four, three and two chords respectively.} 
			\label{PIC_chorddiagrams}
		\end{center}
	\end{figure}
	
	There is an obvious bijection $D$ between traces of (monomial) Dirac words $\sw$ and chord diagrams that assigns to each vertex a letter (respecting the relative ordering) and represents duplicate letters by chords. The cyclicity of $\tr(\sw) = 2\sym(\sw+\sw^*)$ is manifest in the base cycle of the chord diagram $D(\sw)$ and since it is also symmetric it does not make a difference whether we choose to label the vertices clockwise or anti-clockwise.\\
	
	In order to include products of traces, in particular those that contain contractions of matrices in different traces, the usual definition of chord diagrams is not enough, so we generalise as follows:
	\begin{defi}\textup{\textbf{(Generalised chord diagram)}}\\
		A \textup{generalised chord diagram} of order $n=(n_1,\dotsc, n_{\ell})$ is a graph that consists of $\ell$ chord diagrams $D_i\in \cD^{n_i}_{k_{i}}$. In addition to chords within each base cycle a generalised chord diagram may also contain edges between vertices in different base cycles, which we will also call chords (but each vertex is still at most 3-valent). We write $N=\sum n_i$ for the total order of the diagram and denote with $\cD^{(n_1,\dotsc, n_{\ell})}_k$ the set of chord diagrams with the respective number and size of base cycles and $k\leq N$ chords.
	\end{defi}
	
In the following we will always just write chord diagram for the general version. Finally, for the discussion below we need to sort the edges of a chord diagram into three distinct sets, which we do via colouring.
\begin{defi}\textup{\textbf{(Edge $k$-colouring)}}\label{def_edge_colouring}\\
		Let $G$ be a graph and $K$ a finite set consisting of $k$ \textup{colours}. Then a map $\kappa: E_G\to K$ is called a $k$\textup{-edge-coloring} if for every vertex $v$ of $G$ all edges incident to it are assigned different colours, i.e. if $\kappa$ is injective on $\partial^{-1}(v\times V_G)\subset E_G$ for all $v\in V_G$.
	\end{defi}
	
	The number of colours needed to colour a given graph is given by Vizing's theorem to be either the maximal degree $\Delta$ of the graph or $\Delta+1$ \cite{Vizing_1964_chromNumber}. Clearly, each chord diagram $D$ admits an edge 3-coloring $\kappa: E_D \to \{0,1,2\}$ - sometimes called \textit{Tait colouring} \cite{Tait_1880_colourings} - where two alternating colours $1$ and $2$ are assigned to the edges of the base cycles and the third colour $0$ to all chords. Fix one of the $2^{\ell}$ possibilities of such a colouring. This edge colouring induces a unique (up to permutations of colours) double cover $\{ E_D^{01}, E_D^{02}, E_D^{12}\}$ of the chord diagram in which the components $E_D^{ij} = \kappa^{-1}(\{i,j\})$ are given by edge subsets that have exactly 2 different colours and we write $E_D^0$, $E_D^1$ and $E_D^2$ for the respective single colour edge subsets. Furthermore, each two-coloured edge subset can be decomposed into collections $\cC_D^{ij}, \cP_D^{ij}$ of cycles and paths with $\cP_D^{12} = \emptyset$ and $|\cC_D^{12}| = \ell$ since the bases are the only cycles with these two colours. The two-coloured paths between the 2-valent vertices of $D$ can always be combined to form three-coloured cycles by joining all paths in their shared initial or final vertices. Contracting each path in $\cP_D^{01}$ and $\cP_D^{02}$ to a single edge of colour $1$ or $2$ projects the three-coloured cycles onto a generalised chord diagram $D'$ that consists of a disjoint union of base cycles without any chords. Specifically, it defines a map
	\al{
		\pi_0 : \cD_k^n \to \cD_0^{n'}
	}
	with $n' = (n_1', \dotsc, n_{\ell'}')$, $N' = N-k\leq N$ and $\max\{0,\ell-k\} \leq \ell' \leq \ell+k$. The number of two-coloured and three-coloured cycles is the central combinatorial property that we will need later, so we introduce a separate notation for it:
	\al{
		c_2(D) \defeq |\cC_D^{01}| + |\cC_D^{02}| + \ell\hspace{2cm}		c_3(D)  \defeq c_2(D') = \ell'
	}
	From now on we often abbreviate two-coloured cycle and three-coloured cycle as 2-cycle and 3-cycle respectively.
	
	\begin{figure}[h]
		\begin{center}
			\begin{subfigure}{0.9\textwidth}
				\includegraphics{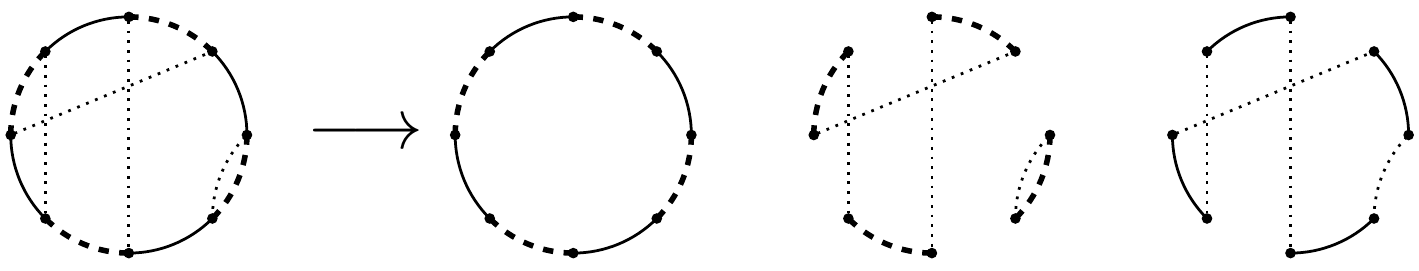}
				\caption{There are no free vertices, so there are no paths but a total of 4 two-coloured cycles.}\vspace{3mm}
				\label{PIC_doublecovers_a}
			\end{subfigure}
			\begin{subfigure}{0.9\textwidth}
				\includegraphics{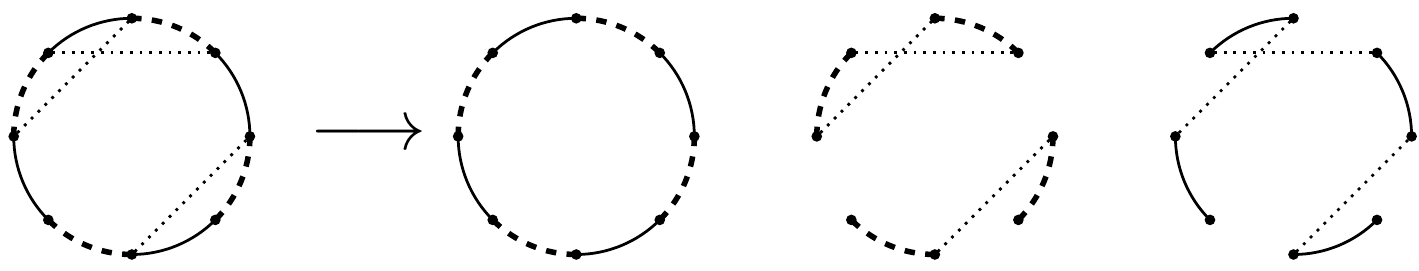}
				\caption{The $01$-component consists of a single path while the $02$-component contains a path and a 2-cycle.}\vspace{3mm}
				\label{PIC_doublecovers_b}
			\end{subfigure}
			\begin{subfigure}{0.9\textwidth}
				\includegraphics{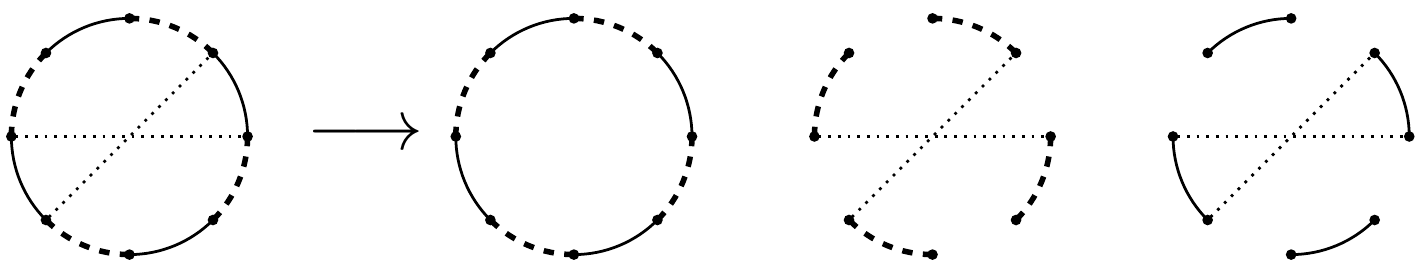}
				\caption{There are 4 paths in total, but they all combine to form a single three-coloured cycle.}
				\label{PIC_doublecovers_c}
			\end{subfigure}
		\end{center}
		\caption{Cycle double covers of the chord diagrams from \reffig{PIC_chorddiagrams}.}
		\label{PIC_doublecovers}
	\end{figure}
	\begin{exam}\label{example_three_colour_stuff}
		In drawings we use different line types to represent the colours:
		\al[*]{
			0 \sim \dotted \hspace{2cm} 1 \sim \scalar \hspace{2cm} 2 \sim \dashed
		}
		Let $D_1, D_2, D_3$ be the three chord diagrams from \reffig{PIC_chorddiagrams}, from left to right. Their colourings and two-coloured components are depicted in \reffig{PIC_doublecovers}. For $D_1$ there are no free vertices, so all two-coloured components are cycles and 
		\al[*]{
			c_3(D_1) = 0 \qquad c_2(D_1) =  1 + 2 + 1 = 4.
		}
		$D_2$ contains two free vertices - 4 and 6 - with two different two-coloured paths between them, forming a three-coloured cycle. Overall $c_2(D_2) = 2$ and $c_3(D_2)=1$. Finally, $D_3$ has four free vertices. There are two 2-cycles, the base and one other coloured $\{\dotted, \dashed\}$. The four paths form a single 3-cycle, so $c_2(D_3) = 2$ and $c_3(D_3)=1$.
	\end{exam}

	\begin{figure}[h]
		\begin{center}
			\includegraphics[width=0.7\textwidth]{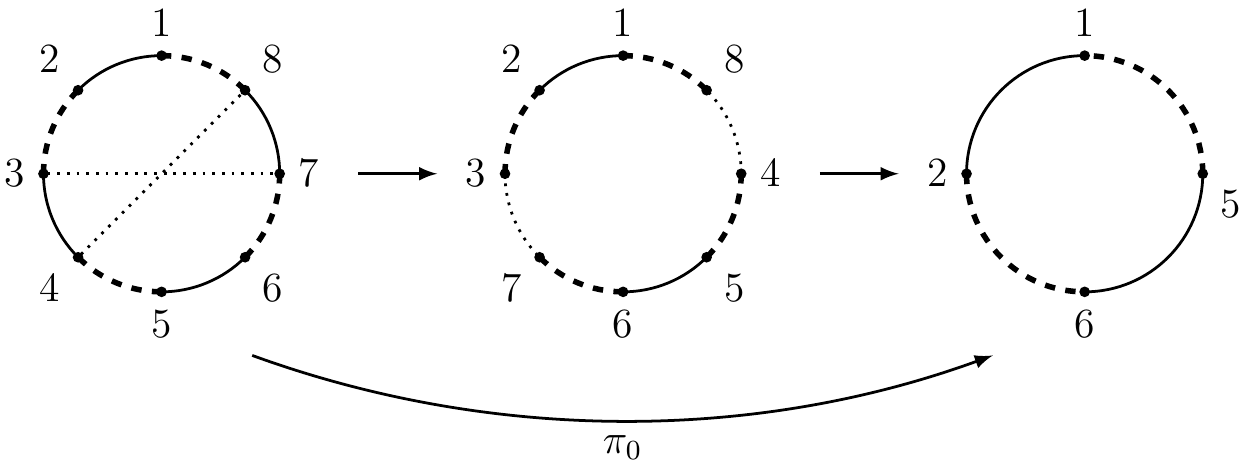}
			\caption{Visualisation of the projection map $\pi_0$ for the case of diagram $D_3$ from \reffig{PIC_doublecovers_c}. } 
			\label{PIC_projection}
		\end{center}
	\end{figure}

Understanding the cycle structure of chord diagrams and how it changes upon addition of more chords is the main task ahead.
\begin{prop}\label{prop_addchord_cases_cycles}
	Let $D_0 \in \cD_{k-1}^n$ with $1\leq k\leq N = \sum n_i$ and $D\in \cD_k^n$ result from $D_0$ by adding a chord between two vertices $i$ and $j$. Then there are the following possibilities.
	\begin{enumerate}
		\item If $i$ and $j$ are in the same 3-cycle and
		\begin{enumerate}
			\item both segments between them consist of a single path, then
			\al{
				c_2(D) = c_2(D_0)+2 \qquad c_3(D) = c_3(D_0)-1.
			}
			\item one segment consists of a single path and the other of a (necessarily odd) number of paths larger than 1, then
			\al{
				c_2(D) = c_2(D_0)+1 \qquad c_3(D) = c_3(D_0).
			}
			\item both segments between them consist of a nonzero even number of paths, then
			\al{
				c_2(D) = c_2(D_0) \qquad c_3(D) = c_3(D_0).
			}
			\item both segments between them consist of an odd ($\geq 3$) number of paths, then
			\al{
				c_2(D) = c_2(D_0) \qquad c_3(D) = c_3(D_0)+1.
			}
		\end{enumerate}
		\item If $i$ and $j$ are in different 3-cycles, then
			\al{
				c_2(D) = c_2(D_0) \qquad c_3(D) = c_3(D_0)-1.
			}
	\end{enumerate}
	\begin{proof} 
		The cases $1. (a)$ and $1. (b)$ are apparent since any single path is completed by a chord to form a new 2-cycle, while the other segment remains a 3-cycle with the chord in place of the former path. In $1.(c)$ both segments have two different coloured edges on their ends and their opposing colour ends are incident to each other in $i$ and $j$. Hence the new chord bridges the equally coloured endings which results in a new 3-cycle. Visually, a plane 3-cycle is twisted into an $\infty$-shape, or alternatively, one segment is cut out, flipped and glued back into the 3-cycle with chords as glue. In $1.(d)$ both ends of either segment have the same colour, such that the new chord cleanly separates the 3-cycle into two new 3-cycles. Finally, in the second case the edges of either colour incident to $i$ are connected by the chord to the equally coloured edge incident to $j$ in the other 3-cycle such that a single new cycle results.
	\end{proof}
\end{prop}	

	\begin{exam}\label{example_chord_add_prop}
		Adding a chord between the two free vertices of $D_2$ from example \ref{example_three_colour_stuff} (cf. \reffig{PIC_doublecovers_b}) falls into case $1.(a)$. All six possible ways to add a chord between any two vertices of $D_3$ (\reffig{PIC_doublecovers_c}) are examples of case $1.(b)$. To illustrate the other cases one needs either larger very complicated diagrams or almost trivial cases, so for simplicity consider $D_0\in \cD^3_0$ to be the empty chord diagram of order $3$, in which every single edge is a path:
	\begin{center}
		\includegraphics[width=0.7\textwidth]{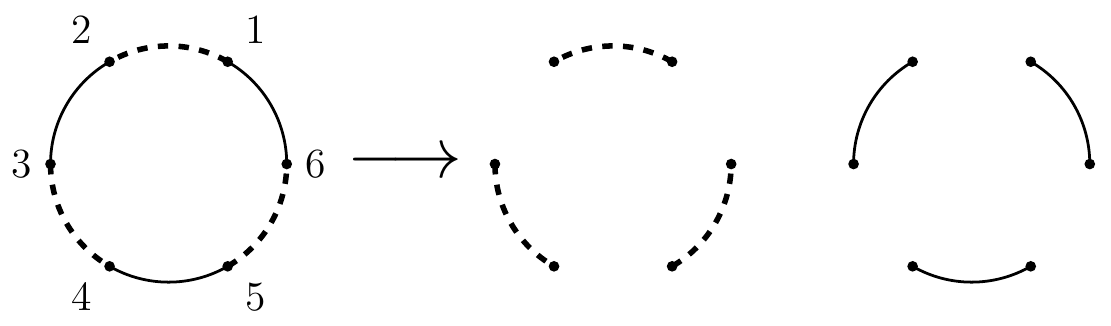}
	\end{center}
	New chords between any pair of vertices separated by one other vertex (say $(1,3)$) correspond to case $1.(c)$. Adding a chord between any of the pairs $(1,4)$, $(2,5)$ or $(3,6)$ corresponds to case $1.(d)$, where the base cycle is split into two new three-coloured cycles. Say the chord $(1,4)$ is added. Then additionally connecting $(2,5)$ or $(3,6)$ would be examples for case $2$.
	\end{exam}

\subsection{Cycle words and diagram contraction}

	For this section we only consider the single base cycle case $\ell=1$. The results are then generalised in the following section. Above we already mentioned the relation between traces of monomial Dirac words and chord diagrams. Let $\sw \in \sD$ be a Dirac word such that $D(\sw) \in \cD^n_k$ for $k<n$. Then $D(\sw)$ contains at least one 3-cycle and $2(n-k)$  2-valent vertices, corresponding to the non-duplicate letters of $\sw$. The structure of $D$ then tells us how to arrange these letters into new words in $\bar\sw(D) \in \bar\sD$ which will allow us to compute the contractions of duplicate letters easily.

	\begin{defi}\textbf{\textup{(Cycle words)}}\label{def_cycle_words}\\
		Let $D\in \cD^n_k$ be a chord diagram with the canonical edge 3-colouring introduced above and $D'=\sqcup_{i=1}^{\ell'} D'_i= \pi_0(D)$. Then for each $D_i'$ consider the words $\su_i \in \bar\sD$ that satisfy $D(\su_i) = D_i'$. Up to cyclic shifts there are four such words for each $D_i'$ and they are related to each other as $\su_i, \tilde\su_i, \su_i^*$ and $\widetilde{(\su_i^*)}$. Using these words we define the cycle word associated to $D$ as
		\al{
			\bar\sw(D) \defeq \frac{1}{2} \Bigg(\prod_{i=1}^{\ell'} \sym(\su_i) + \prod_{i=1}^{\ell'} \sym(\su_i^*)\Bigg).
		}
	\end{defi}
	
	\begin{exam}\label{example_cycleWord}
		Consider the chord diagram $D_3$ from \reffig{PIC_projection} and \reffig{PIC_doublecovers_c}, previously discussed in examples \ref{example_three_colour_stuff} and \ref{example_chord_add_prop}. It has the four free vertices $1,2,5$ and $6$, with four paths $1-2$, $2-3-7-6$, $6-5$ and $5-4-8-1$ combining to one three-coloured cycle. Note that after projection to a base cycle the free vertices are not in the original order anymore. Choose for example $\su = \sa_1\sa_2\sa_6\sa_5$. Then 
		\al{
			\bar \sw(D_3) = \frac{1}{4}\Big( \sa_1\sa_2\sa_6\sa_5 + \sa_5\sa_6\sa_2\sa_1 + \sa_2\sa_6\sa_5\sa_1 + \sa_1\sa_5\sa_6\sa_2\Big)
		}
		
		 For an example with multiple cycles consider the empty order 3 diagram also discussed in example \ref{example_chord_add_prop} together with a single chord between $(3,6)$. One cycle consists of the two paths $2-3-6-1$ and $1-2$ which gives the word $\su_{1} = \sa_2\sa_1$ while the other in analogous fashion gives $\su_{2}=\sa_4\sa_5$. The cycle word is then
		 \al{
		 	\frac{1}{8}\Big( ( \sa_2\sa_1 + \sa_1\sa_2 )( \sa_4\sa_5 + \sa_5\sa_4 ) + ( \sa_1\sa_2 + \sa_2\sa_1 )( \sa_5\sa_4 + \sa_4\sa_5 ) \Big)
			= \frac{1}{4}\Big( ( \sa_2\sa_1 + \sa_1\sa_2 )( \sa_4\sa_5 + \sa_5\sa_4 ) \Big).
		 }
	\end{exam}
	
	In the case $k=0$ one has $\pi_0(D(\sw)) = D(\sw) \in \cD^n_0$ with $\ell'=\ell=1$ and therefore
	\al{
			\bar\sw(D(\sw))= \frac{1}{2} \Big(\sym(\sw) +  \sym(\sw^*)\Big) = \frac{1}{4}\tr(\sw)
	}
	by theorem \ref{theo_wordtrace}. This is quite sensible since we can interpret the ``contraction'' of a word without duplicate letters to contract as the expansion into $\delta_{ij}$ via the trace recursion formula, divided by $4=\tr(1)$. On the other hand one sees that if $k=n$ then there are no more 3-cycles in $D(\sw)$ and $\bar\sw(D(\sw))=1$. More generally we find the following relation between $\sw$ and $\bar \sw$.
	\begin{theo}\label{theo_contraction_diagrams}
		Let $\sw \in \sD$ be a monomial Dirac word such that the associated chord diagram $D(\sw) \in \cD^n_k,\ 0 \leq k\leq n$. Then
		\al{
			\tr(\sw) = (-2)^{k+c_2(D(\sw))+c_3(D(\sw))}  \bar \sw(D(\sw)).
		}
	\begin{proof}
		As discussed above the case $k=0$ gives $\tr(\sw) = 4\bar\sw(D(\sw))$ where $4 = (-2)^{0+1+1}$ such that the claim holds for all $n$. For $0<k\leq n$ we prove by induction over the number of chords. Let $\sw'\in\sD$ such that $\sw = \delta_{ij}\sw'$. We abbreviate $c_2\equiv c_2(D(\sw))$, $c_3\equiv c_3(D(\sw))$ and $\bar \sw \equiv \bar \sw(D(\sw))$, and write $c_2'$, $c_3'$ and $\bar\sw'$ for the corresponding objects resulting from $\sw'$. $D(\sw)$ is $D(\sw')$ together with a chord between vertices $i$ and $j$ and we need to consider the same five cases as in proposition \ref{prop_addchord_cases_cycles}. The idea is the same for all cases: Use the contraction relation \refeq{eq_contraction_words_elegant} to compute $\delta_{ij}\bar\sw' = N \bar\sw$ where $N$ is some integer factor. Then confirm that both the change in term structure of the cycle word and the new integer factor is in accordance with change in cycle structure and cycle numbers $c_2$ and $c_3$ as discussed in proposition \ref{prop_addchord_cases_cycles}.
		\begin{enumerate}
		\item If $i$ and $j$ are in the same 3-cycle (base cycle of $\pi_0(D(\sw'))$) with word $\su_{ij}$ and
		\begin{enumerate}
			\item both segments between them consist of a single path, then $\sym(\su_{ij}) = \frac{1}{2}(\sa_i\sa_j+\sa_j\sa_i) = \sym(\su_{ij}^*)$ such that $\delta_{ij}\bar\sw' = 4 \bar \sw$, where $\bar \sw$ is the same as $\bar\sw'$ except that the entire base cycle that contained $i$ and $j$ -- and no other vertices -- has been removed from both products. Hence, in accordance with $\ref{prop_addchord_cases_cycles}$ we have $c_3 = c_3'-1$. Furthermore we have $c_2 = c_2'+2$ and one more chord ($k-1 \to k$) such that $4=(-2)^{1+2-1}$ and
				\al{\nonumber
					\tr(\sw) = \delta_{ij} \tr(\sw') &= (-2)^{k-1+c_2'+c_3'} \delta_{ij}\bar\sw'\\\nonumber
						&=  (-2)^{k-1+c_2'+c_3'} 4\bar\sw\\
						&=  (-2)^{k+c_2+c_3}\bar\sw.
				}
			\item one segment consists of a single path and the other of a (necessarily odd) number of paths larger than 1, then $\su_{ij} = \sv_1\sa_i\sa_j\sv_2$ for some words $\sv_1,\sv_2\in \bar \sD$. Multiplying with $\delta_{ij}$ extracts the factor $4=\sa_i^2$ but otherwise leaves the product structure of $\bar\sw'$ intact (and in particular $c_3=c_3'$). There is one additional chord and one new 2-cycle, absorbing the factor $4=(-2)^{1+1+0}$:
				\al{
					\tr(\sw) =  (-2)^{k-1+c_2'+c_3'} 4\bar\sw = (-2)^{k+c_2+c_3}\bar\sw
				}
			\item both segments between them consist of a nonzero even number of paths, then $\su_{ij} = \sv_1\sa_i\sv_2\sa_j\sv_3$ for some words $\sv_1,\sv_2, \sv_3\in \bar \sD$ with $|\sv_2|$ odd. One can use
				\al{
				\delta_{ij}\sa_i\sv_2\sa_j = \sa_i\sv_2\sa_i = -2\tilde\sv_2
				}
				and as before one finds (with $c_2 = c_2'$ and $c_3=c_3'$)
				\al{
					\tr(\sw) =  (-2)^{k-1+c_2'+c_3'} (-2)\bar\sw = (-2)^{k+c_2+c_3}\bar\sw.
				}			
			\item both segments between them consist of an odd number of paths (at least three each), then $\su_{ij} = \sv_1\sa_i\sv_2\sa_j\sv_3$ for some words $\sv_1,\sv_2, \sv_3\in \bar \sD$ with $|\sv_2|$ even. One can use proposition \ref{prop_commute_words} to find
				\al{
				\delta_{ij}\su_{ij} =  \sv_1\sa_i\sv_2\sa_i\sv_3 = 4\sv_1\sym(\sv_2^*)\sv_3 = 4\sv_1\sv_3\begin{cases}
						 \sym(\sv_2^*) &\text{ if $|\sv_1|, |\sv_3|$ even}\\[1mm]
						 \sym(\sv_2)&\text{ if $|\sv_1|, |\sv_3|$ odd}
					 \end{cases}
				}
				and write
				\al{\nonumber
					\sym(\delta_{ij}\su_{ij}) &= 2  \begin{cases}
						 \sv_1\sv_3\sym(\sv_2^*) + \sym(\sv_2^*)\tilde\sv_3\tilde\sv_1 &\text{ if $|\sv_1|, |\sv_3|$ even}\\[1mm]
						 \sv_1\sv_3\sym(\sv_2) + \sym(\sv_2)\tilde\sv_3\tilde\sv_1 &\text{ if $|\sv_1|, |\sv_3|$ odd}
					 \end{cases}\\[2mm]
					 & = 4 \sym(\sv_1\sv_3)\begin{cases}
						 \sym(\sv_2^*) &\text{ if $|\sv_1|, |\sv_3|$ even}\\[1mm]
						 \sym(\sv_2)  &\text{ if $|\sv_1|, |\sv_3|$ odd}.
					\end{cases}
				}
				If the even (odd) case applies to $\su_{ij}$ then the odd (even) case can be used to find the analogous result (with $(\sv_1\sv_3)^*$) for $\su_{ij}^*$. We see the expected splitting into two 3-cycles realised in the products. One finds altogether
				\al{
					\tr(\sw) =  (-2)^{k-1+c_2'+c_3'} 4\bar\sw = (-2)^{k+c_2+c_3}\bar\sw.
				}
		\end{enumerate}
		\item If $i$ and $j$ are in different 3-cycles, then we need to consider a product $\sym(\su_i)\sym(\su_j)$. One can always choose representatives $\su_i$ and $\su_j$ such that $\sa_i$ and $\sa_j$ are either their first or last letter respectively. Hence, there exist words $\sv_1,\sv_2 \in \bar\sD$ such that
		\al{
			\delta_{ij}\sym(\su_i)\sym(\su_j) & = \frac{1}{4}( \sa_i\sv_1 + \tilde\sv_1\sa_i)( \sa_i\sv_2 + \tilde\sv_2\sa_i)
											= \sym(\tilde\sv_2\sv_1).
		}
		The factor here is $1=(-2)^{1+0-1}$ where we have one more chord but lost one 3-cycle, so here, too, everything works out as claimed.
	\end{enumerate}
	\end{proof}
	\end{theo}
	
	In the case of a single trace discussed here one could have simply used $\prod \sym(\su_i)$ as cycle word and found the same result. However, next we want to consider products which have duplicate letters within different traces. In those cases it is crucial to consider the full $\bar\sw(D(\sw))$ with products over both $\sym(\su_i)$ and $\sym(\su_i^*)$ as defined above.

	\begin{rem}\label{REM_oddwords}
		Above we only discussed contraction of traces of even words. In practice one would also like to contract odd words, which are associated to ``open'' fermion lines in a Feynman graph. For contraction of such a word $\sw'\in \sD$ with $|\sw'|$ odd consider the word $\sw=\sw_1\sa_i\sw_2$ where $\sa_i\in\sA$ is a dummy letter that does not occur in $\sw'$, $\sw_1\sw_2 = \sw'$ and $\sw_2$ starts with the first letter that occurs only once in the $\sw'$, i.e. is not contracted. The trace and its contraction of $\sw$ can be computed as above. The contraction of the odd word is then simply obtained by dividing by 4 and ``unsymmetrising'' the factor corresponding to the  3-cycle that contains the dummy vertex in the two products in the cycle word, i.e.
		\al[*]{
			\sym(\su_i)  \to \su_i \text{ or } \tilde\su_i \qquad \text{ and } \qquad 
			\sym(\su_i^*) \to \su_i^* \text{ or } \tilde\su_i^*
		}
		where the choice is fixed by demanding that upon evaluation $\sa_i\to 1$ the first letter of the unsymmetrised word is the first letter of $\sw_2$. See also example \ref{example_newContr} below.
	\end{rem}
	
	\begin{exam}\label{example_newContr}
		We return again to the contraction of $g_{\nu_2\nu_4}g_{\mu_2\mu_4} \gamma_{\Gamma_1}$ from example \ref{EX_oldContr}, corresponding to the Dirac word
		\al{
			\sw = \delta_{37}\delta_{48}\sa_1\sa_2\sa_3\sa_4\sa_5\sa_6\sa_7\sa_8
		}
		whose chord diagram is again $D(\sw)=D_3$, the rightmost diagram in \reffig{PIC_chorddiagrams} which we already discussed in all the previous examples. In example \ref{example_three_colour_stuff} we found $c_2(D_3)=2$ and $c_3(D_3)=1$, and in example \ref{example_cycleWord} we saw
		\al{
			\bar\sw(D_3) = \frac{1}{2}(\sym(\sa_1\sa_2\sa_6\sa_5) + \sym(\sa_2\sa_6\sa_5\sa_1)) = \frac{1}{4}\tr(\sa_1\sa_2\sa_6\sa_5).
		}
		Therefore $\tr(\sw) = (-2)^{2+2+1}\frac{1}{4}\tr(\sa_1\sa_2\sa_6\sa_5) = -8\tr(\sa_1\sa_2\sa_6\sa_5)$, which is the same result as in the previous manual computation.\\[1mm]
		Let $\sv = \sa_2\sa_3\sa_4\sa_5\sa_6\sa_3\sa_4$ be the odd word such that $\sw=\sa_1\sv$. We compute its contraction following remark \ref{REM_oddwords}. There is only one factor in the cycle word to be unsymmetrised and the choice is such that $\sa_2$ is the first letter after removal of $\sa_1$. One finds
		\al{
			\sv = (-2)^5\frac{1}{4}\frac{1}{2}(\sa_1\sa_2\sa_6\sa_5 + \sa_2\sa_6\sa_5\sa_1)_{\sa_1\to 1} = -8\sa_2\sa_6\sa_5,
		}
		which is the expected result.
	\end{exam}
	
	\begin{exam}
		We can compute a larger example, like the contraction of an 18 letter word, to demonstrate the efficiency of this contraction formalism. Let
		\al{\nonumber
			\tr(\sw) &= \tr(\sa_1\sa_2\sa_3\sa_4\sa_1\sa_6\sa_2\sa_8\sa_9\sa_{10}\sa_9\sa_3\sa_{10}\sa_{14}\sa_4\sa_8\sa_6\sa_{14})\\\nonumber
				& = (-2)^3\tr(\sa_4\sa_3\sa_2\sa_6\sa_2\sa_8\sa_{10}\sa_3\sa_{10}\sa_6\sa_8\sa_4)\\\nonumber
				& = (-2)^5\tr(\sa_4\sa_3\sa_6\sa_8\sa_3\sa_6\sa_8\sa_4)\\ \nonumber
				& = (-2)^7\tr(\sa_4\sa_6\sa_6\sa_4)\\
				& = (-2)^{13}
		}
		where we already combined multiple contractions in the same line and chose an efficient order of contractions. For our formalism we simply count the number chords ($k=9$), 3-cycles ($c_3(D(\sw))=0$) and 2-cycles ($c_2(D(\sw))=4$, the base, two depicted below on the left and one below on the right). Hence we have indeed $\tr(\sw) = (-2)^{9+4+0}1 = (-2)^{13}$.
		\begin{center}
			\includegraphics[width=0.75\textwidth]{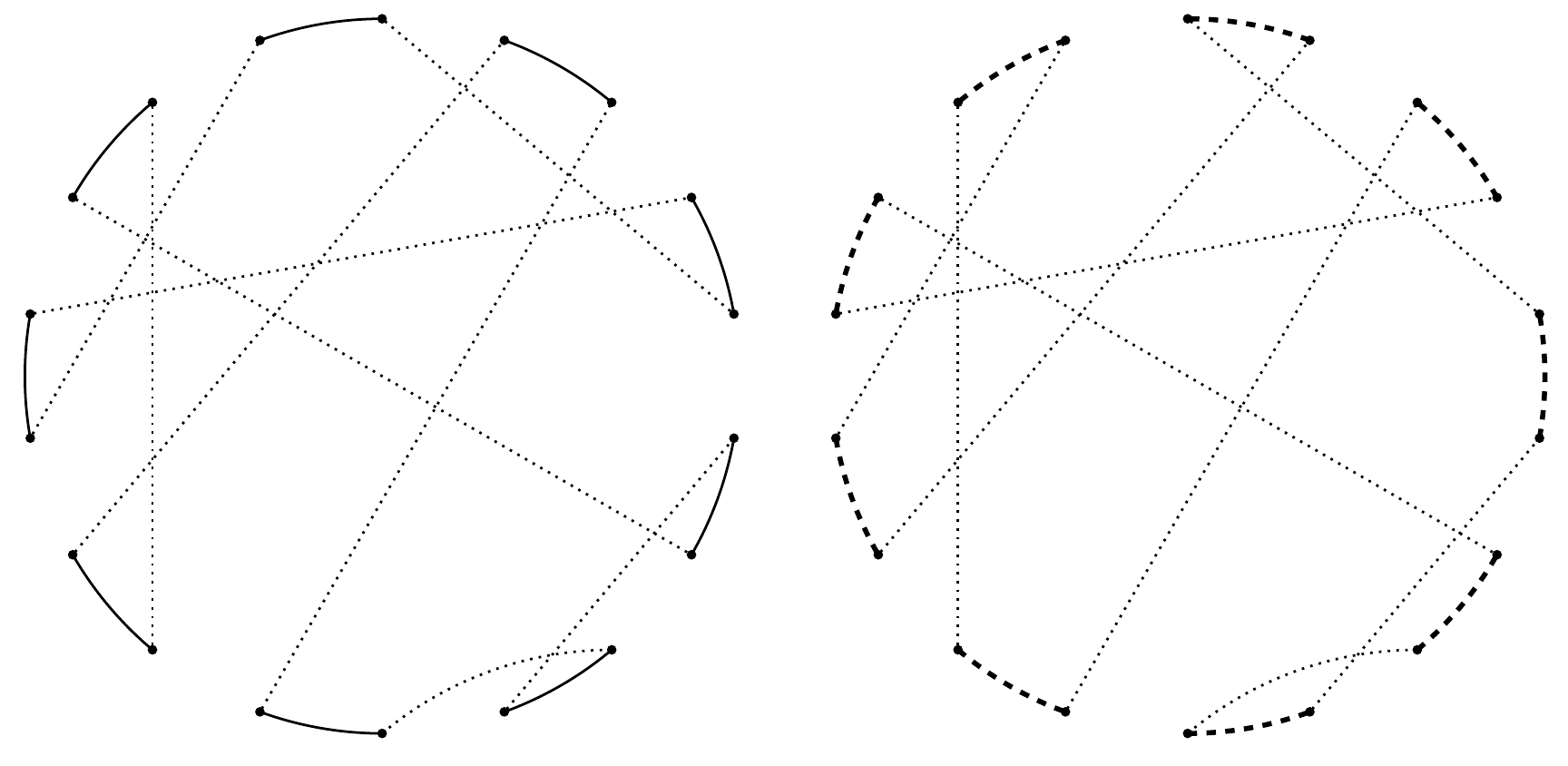}
		\end{center}
	\end{exam}

\subsection{Multiple traces}
	Above we considered contraction of single traces but theorem \ref{theo_contraction_diagrams} can be generalised to arbitrary products of traces - including contraction of letters occurring in different traces - without much effort.\\
	
	Consider first two words $\sw_1,\sw_2\in\sD$ without any shared letters and $D_i=D(\sw_i)\in\cD^{n_i}_{k_i}$ for $i=1, 2$ their respective chord diagrams. Multiplying their traces gives
	\al{
		\tr(\sw_1)\tr(\sw_2) = (-2)^{k_1+k_2+c_2(D_1)+c_2(D_2)+c_3(D_1)+c_3(D_2)}  \bar \sw(D_1)\bar \sw(D_2)
	}
	where
	\al{\nonumber
		\bar\sw(D_1)\bar\sw(D_2) &= \frac{1}{4} \Bigg(\prod_{i=1}^{\ell_1'} \sym(\su_{1,i})\prod_{j=1}^{\ell_2'} \sym(\su_{2,j}) 
			+ \prod_{i=1}^{\ell_1'} \sym(\su_{1,i}^*)\prod_{j=1}^{\ell_2'} \sym(\su_{2,j}^*)\\
		& \qquad + \prod_{i=1}^{\ell_1'} \sym(\su_{1,i})\prod_{j=1}^{\ell_2'} \sym(\su_{2,j}^*) 
			+ \prod_{i=1}^{\ell_1'} \sym(\su_{1,i}^*)\prod_{j=1}^{\ell_2'} \sym(\su_{2,j})\Bigg).\label{eq_tracemult}
	}
	Consider the disjoint union of the two chord diagrams $D_{12}= D_1\sqcup D_2$. The terms in \refeq{eq_tracemult} can be interpreted as two different cycle words associated to $D_{12}$, corresponding to two different colourings of the base cycles of $D_1$ and $D_2$.\\
	
	Assuming that all diagrams use the same colour for their chords, there are $2^{\ell}$ possible colourings of the $\ell$ base cycles with the other two colours - visible as four terms in \refeq{eq_tracemult}. Combining the terms pairwise (the two in the upper line and the two in the lower line) one has a sum over the $2^{\ell-1}$ \textit{relative colourings} of the base cycles. Earlier we defined the map $D$ from Dirac words $\sD$ to chord diagrams. Clearly it can be extended to word tuples $(\sw_1,\dotsc,\sw_{\ell})\in\sD^{\ell}$, mapping them to chord diagrams with $\ell$ base cycles as long as the concatenation $\sw_1\dotsm\sw_{\ell}\in\sD$, i.e. as long as no letter appears more than twice in the tuple. As we have seen above there are $2^{\ell-1}$ different relative colourings of the base cycles. Clearly the set of three-coloured cycles and the projection of such a chord diagram then depend on the choice of colouring $c$. Similarly the cycle numbers $c_2(D,c)$ and $c_3(D,c)$ depend now on the choice of colouring. Using this we can simply extend the cycle word definition \ref{def_cycle_words} to a generalised chord diagram together with a particular colouring as
	\al{
			\bar\sw(D,c) \defeq \frac{1}{2} \Bigg(\prod_{i=1}^{\ell'} \sym(\su_i) + \prod_{i=1}^{\ell'} \sym(\su_i^*)\Bigg).
	}
	where of course $\ell'$ and all the $\su_i$ are now also colour dependent via $\pi_0^c$. The overall cycle word for a generalised chord diagram $D\in\cD_{k}^{(n_1,\dotsc,n_{\ell})}$ is then
	\al{\nonumber
		\bar\sw(D) &\defeq \frac{1}{2^{l-1}}\sum_c (-2)^{c_2(D,c)+c_3(D,c)} \bar\sw(D,c)
	}
	Every summand is of the same form as the cycle word for a single chord diagram, so theorem \ref{theo_contraction_diagrams} can be applied term by term and extends fully to generalised chord diagrams. In particular the addition of a chord between different base cycles (corresponding to contraction of letters in different traces) can be treated the same as case 2 (new chord between different 3-cycles) in the proof. 

	\begin{figure}[h]
		\begin{center}
			\begin{subfigure}{0.95\textwidth}
				\includegraphics{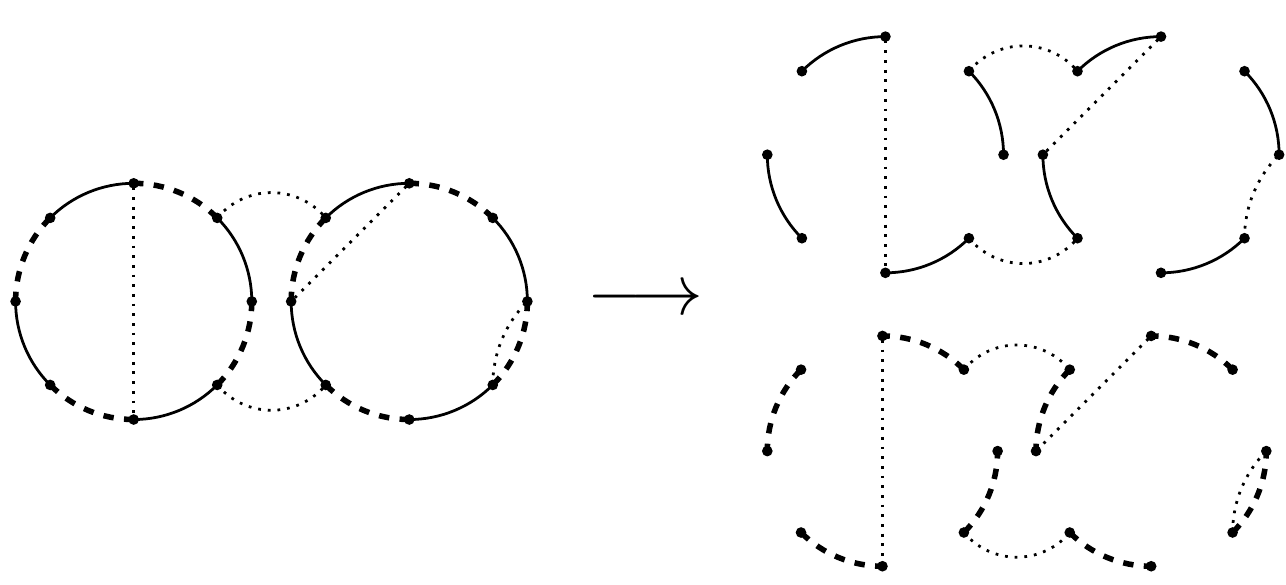}
			\end{subfigure}
			\hrule
			\begin{subfigure}{0.95\textwidth}
				\includegraphics{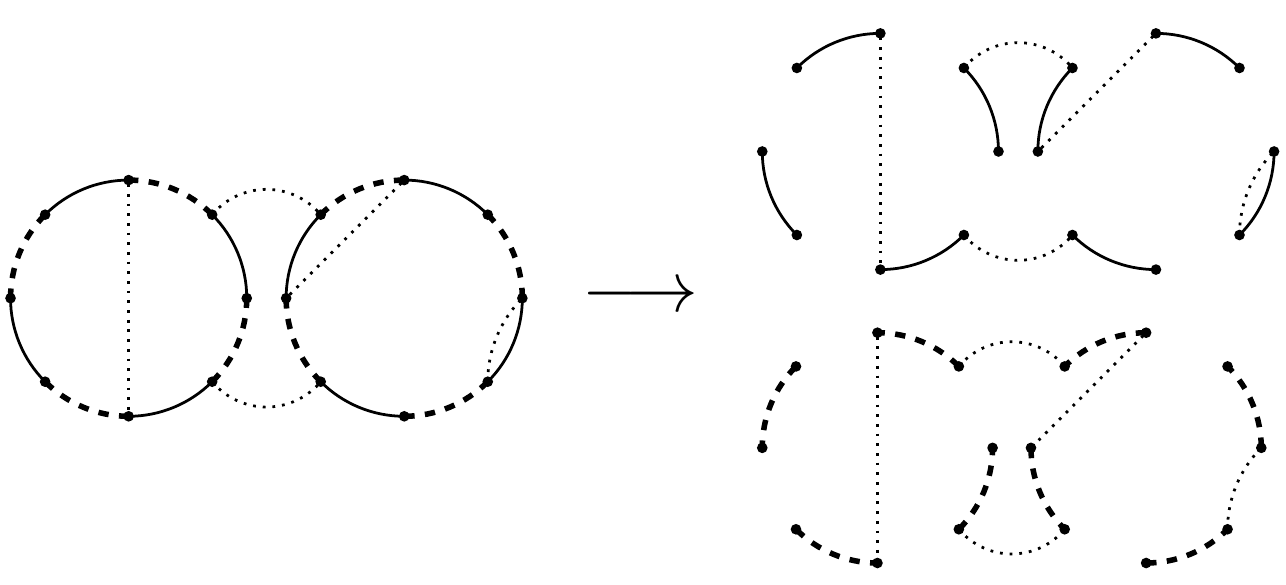}
			\end{subfigure}
		\end{center}
		\caption{The two-coloured subgraphs of the same chord diagram for two different relative colourings. One can see the different structures that result in different cycle words and numbers.}
		\label{PIC_example_multi}
	\end{figure}

	\begin{cor}\label{cor_generalisation}
		Let $(\sw_1,\dotsc,\sw_{\ell}) \in \sD^{\ell}$ be a tuple of Dirac words such that $D \equiv D(\sw_1,\dotsc,\sw_{\ell})\in\cD_{k}^{(n_1,\dotsc,n_{\ell})}$. Then
		\al{\nonumber
			\tr(\sw_1)\dotsm\tr(\sw_{\ell}) &= (-2)^k \bar\sw(D)\\[1mm]\nonumber
								& = \frac{(-2)^k}{2^{\ell-1}}\sum_c (-2)^{c_2(D,c)+c_3(D,c)} \bar\sw(D,c).
		}		
	\end{cor}

	\begin{exam}
		Consider the chord diagram depicted with its two different relative colourings in \reffig{PIC_example_multi}. Labelling counter-clockwise and starting with the uppermost vertex of the left base cycle it corresponds to
		\al{\nonumber
			\delta_{1,5}\delta_{6,12}\delta_{8,10}\delta_{9,11}\delta_{14,15}&
				\tr(\sa_1\sa_2\sa_3\sa_4\sa_5\sa_6\sa_7\sa_8)
				\tr(\sa_9\sa_{10}\sa_{11}\sa_{12}\sa_{13}\sa_{14}\sa_{15}\sa_{16})\\\nonumber
			&= (-2)^4\tr(\sa_4\sa_3\sa_2\sa_6\sa_7\sa_8)
				\tr(\sa_8\sa_6\sa_{13}\sa_{16})\\
			&= -(-2)^6\Big( \tr(\sa_4\sa_3\sa_2\sa_7\sa_{13}\sa_{16}) + \tr(\sa_4\sa_3\sa_2\sa_{13}\sa_{16}\sa_7) \Big).\label{eq_example_multi}
		}
		We have a total of five chords in two base cycles, so $k=5$, $\ell=2$. The two colourings each have only one 3-cycle, with corresponding words
		\al{
			\su_1 = \sa_3\sa_4\sa_{16}\sa_{13}\sa_7\sa_2 \quad\text{ and }\quad \su_2 = \sa_3\sa_4\sa_7\sa_{16}\sa_{13}\sa_2
		}
		respectively, which are up to cyclic shifts and reversal the two words in the traces in \refeq{eq_example_multi}. Both have three 2-cycles, the two bases and one between vertices $14$ and $15$. Therefore we compute
		\al{\nonumber
			\frac{(-2)^5}{4} &\Bigg( 
				(-2)^{3+1}\Big(\sym(\su_1) + \sym(\su_1^*)\Big) +
				(-2)^{3+1}\Big(\sym(\su_2) + \sym(\su_2^*)\Big) \Bigg)\\[2mm]
			&= -(-2)^6\Big( \tr( \su_1 ) + \tr( \su_2 ) \Big).
		}

	\end{exam}

\section{Feynman integrals and summation of traces}\label{sec_applic}
	
	We can now go back to the integrand of parametric Feynman integrals and combine the term containing traces of products of Dirac matrices with the metric tensors found in \refeq{eq_qed_int}. For simplicity we restrict the discussion to single fermion loops and graphs of photon propagator type, but the following holds true in general. For multiple fermion loops one simply has to use the more cumbersome notation just introduced in the previous section while fermion propagators, vertex graphs etc. can be treated by introducing a dummy vertex to close the fermion loop and then following remark \ref{REM_oddwords} to make some minor changes to the factors. The overall structure is the same. Finally, in order to avoid further lengthy discussion of notation involving the polynomials $\cp{i}{j}$ in the general gauge case \cite{Golz_2017_CyclePol}, we restrict ourselves to Feynman gauge.\\

	Let $n = (|E_{\Gamma}^{(f)}| + |V_{\Gamma}|)/2$ and label the $2n$ vertices of a chord diagram base cycle with fermion edges and vertices of $\Gamma$, respecting their ordering within the fermion cycle. Working in Feynman gauge we now let $D_0$ be that base cycle together with $n/2$ chords fixed in place between the vertices labelled by vertices of $\Gamma$, such that each chord corresponds to a photon edge of $\Gamma$, including an edge between the external vertices $(i_0,j_0)$. We can then reinterpret the sum in \refeq{eq_qed_int} as a sum over all possible chord diagrams containing $D_0$ and write
	\al{
		N_{\Gamma} \sim \sum_{k=\frac{n}{2}+1}^{n} \sum_{\substack{D\in \cD_k^n\\D\supseteq D_0}}
				\Bigg(\prod_{(i,j)\in E_D^0\setminus E_{D_0}^0} g_{\mu_{i}\mu_{j}}\frac{\cp{i}{j}}{2\kp}\Bigg)
				\prod_{l\in V_D^{(2)} }\frac{\Chi_{\Gamma}^{l,\mu_l}}{\kp}.
	}
	where $E^0_D\subset E_D$ are the chords and $V_D^{(2)} \subset V_D$ are the 2-valent vertices of $D$. Combining this with the Dirac matrices $\gamma_{\Gamma}$ and applying theorem \ref{theo_contraction_diagrams} gives
	\al{\nonumber
		\gamma_{\Gamma}N_{\Gamma} 
				&\sim \frac{g_{\mu_{i_0}\mu_{j_0}}}{4}\sum_{k=\frac{n}{2}+1}^{n} \sum_{\substack{D\in \cD_k^n\\D\supseteq D_0}} (-2)^{ k+c_2(D)+c_3(D)}
				\Bigg( \prod_{(i,j)\in E_D^0\setminus E_{D_0}^0} \frac{\cp{i}{j}}{2\kp}\Bigg)
				\frac{\bar \sw(D,\Gamma)}{\kp^{2n-2k}}\\[2mm]
				&= \frac{g_{\mu_{i_0}\mu_{j_0}}}{2}\sum_{k=\frac{n}{2}+1}^{n} \frac{(-1)^k}{\kp^{2n-k-1}}
				\sum_{\substack{D\in \cD_k^n\\D\supseteq D_0}} (-2)^{ c_2(D)+c_3(D)}
				\Bigg( \prod_{(i,j)\in E_D^0\setminus E_{D_0}^0} \cp{i}{j}\Bigg) \bar \sw(D,\Gamma),\label{eq_num}
	}
	where $\bar\sw(D,\Gamma) = \bar\sw(D)|_{\sa_i \to \gamma_{\mu_i}\Chi_{\Gamma}^{i,\mu_i}}$.\\
	
	Example calculations suggest that by exploiting various identities for graph polynomials it should be possible to sum over all chord diagrams for each $k$, finding an even simpler result of the form $\sum A_i/\kp^i$, with each $A_i$ being expressible as a relatively simple polynomial. However, the combinatorics seem to be quite complex and need to be investigated in depth in future work. What we can do for now, as a preparation and to motivate such a summation, is consider the much simpler summation of the traces without polynomials, i.e. sums of the form $\sum_D (-2)^{s(D)}$ where the $D$ are generalised chord diagrams and $s(D) = c_2(D)+c_3(D)$ is the total number of cycles. Based on these results we then conjecture an expression for the sum $\sum_D (-2)^{s(D)}\prod \cp{i}{j}$ for the case $D\in \cD^n_n$ and argue why that should be enough to completely reduce
\refeq{eq_num} to its simplest possible form.
	
\subsection{Summation without polynomials}
\begin{theo}\label{theo_sumTrace}
	Let $D_0 \in \cD_{k-1}^n$ with $n=(n_1,\dotsc, n_{\ell})$ and $1\leq k\leq \sum n_i$. Define $\bar{\cD}(D_0) \subseteq \cD_k^n$ to be the set
	\al{
		\bar{\cD}(D_0) \defeq \{ D \in \cD_k^n \ | \ D_0\subset D \}
	}
	which contains all chord diagrams that result from $D_0$ by adding a chord in all possible ways. Furthermore, denote with $m=n-k+1$ the number of missing chords in $D_0$. Then
	\al{
		\sum_{ D\in \bar{\cD}(D_0) } (-2)^{s(D)} = -m(m+1)\cdot (-2)^{s(D_0)}.
	}
	\begin{proof}
		Since we only care about the change in total number of cycles the five cases of proposition \ref{prop_addchord_cases_cycles} can be collected into only three cases here: Adding a chord between vertices in
		\begin{enumerate}
			\item the same 3-cycle, separated by an odd number of segments: $s(D) = s(D_0) +1$
			\item the same 3-cycle, separated by an even number of segments: $s(D) = s(D_0)$
			\item different 3-cycles: $s(D) = s(D_0) - 1$
		\end{enumerate}
		We first compute explicitly the cases $k=n$ and $k=n-1$ to illustrate the idea of the proof and then prove for general $k$.\\
		
		For $k=n$ ($m=1$) there are only two free vertices, which are the endpoints of the two paths of a single 3-cycle and there is only one possibility of adding a chord, which belongs to case 1. Hence, 
		\al{
			\sum_{ D\in \bar{\cD}(D_0) } (-2)^{s(D)} = (-2)^{s(D)} = (-2)^{s(D_0)+1} = -2\cdot (-2)^{s(D_0)}.
		}
		
		For $k=n-1$ ($m=2$) there are 6 ways to add a chord and the four free vertices can be arranged either in a single 3-cycle with four paths or two 3-cycles with two paths each. If it is a single 3-cycle, then four of the six ways to add a chord fall into case 1, with odd segments containing one and three paths, respectively. The other way of adding a chord is of case 2, with both segments containing two paths, such that overall
		\al{
			\sum_{ D\in \bar{\cD}(D_0) } (-2)^{s(D)} = 4(-2)^{s(D_0)+1} + 2(-2)^{s(D_0)} = -6 \cdot (-2)^{s(D_0)}.
		}
		If there are two 3-cycles then one has four times case 3 of adding a chord between different cycles and twice case 1 of adding a chord in a 3-cycle with two paths (as in the previous case $k=n$), so one finds the same result
		\al{
			\sum_{ D\in \bar{\cD}(D_0) } (-2)^{s(D)} = 4(-2)^{s(D_0)-1} + 2(-2)^{s(D_0)+1} = -6 \cdot (-2)^{s(D_0)}.
		}
		
		In general there are $\binom{2m}{2}$ possibilities of adding a chord to $D_0$ and the $2m$ free vertices can be partitioned into up to $m$ 3-cycles (base cycles of $\pi_0(D_0)$) as follows
		\al[*]{
			& (2m) \\
			& (2m-2, 2), (2m-4, 4), \dotsc, (\lceil m \rceil, \lfloor m \rfloor ) \\
			& \vdots \\
			& (2, 2, \dotsc, 2)
		}
		Two observations allow us to collect all terms in each of these cases. First, consider a single 3-cycle on $2l$ vertices. Adding a chord separates the cycle into segments of length $l_1$ and $l_2$, $l_1\geq l \geq l_2$. There are $2l$ possibilities for each pair $(l_1,l_2)$ with $l_1 > l > l_2$ and $l$ for $l_1= l = l_2$. By simple counting one finds that this gives $l^2$ instances of case 1 (odd length segments) and $l(l-1)$ of case 2 (even length segments).
		Secondly, consider a set of $\ell$ 3-cycles on $2l_i$ vertices respectively and count only the number of possibilities to add a chord between any two of them. There are $\binom{\ell}{2}$ choices of two cycles, each of which contribute $2l_i\cdot 2l_j$ possibilities to add a chord such that we can express the total number of possibilities as $E_2(2l_1,\dotsc, 2l_{\ell}) = 4E_2(l_1,\dotsc, l_{\ell})$, the evaluation of the elementary symmetric polynomial of degree 2.
		Combining these two results we find that, for a set of $\ell$ 3-cycles on $2m_i$ vertices with $\sum m_i = m$ one has the following number of chord additions corresponding to each case:
		\al[*]{
			\text{case 1} \rightarrow \sum m_i^2\hspace{2cm}
			\text{case 2} \rightarrow \sum m_i(m_i-1)\hspace{2cm}
			\text{case 3} \rightarrow 4E_2(m_1,\dotsc, m_r)
		}
		
		Now it remains to be shown that the sum $\sum_{ D\in \bar{\cD}(D_0) } (-2)^{s(D)}$ yields the same result, regardless of the particular 3-cycle partition present in $D_0$. Assume $D_0$ contains $\ell$ 3-cycles with $m_i$ free vertices such that $(m_1,\dotsc, m_{\ell})$ with $\sum m_i = m$ is the corresponding integer partition of $m\geq 1$. Then
		\al{\nonumber
			\sum_{ D\in \bar{\cD}(D_0) } (-2)^{s(D)} 
			&= (-2)^{s(D_0)}\left( -2\sum_{i=1}^{\ell} m_i^2 +\sum_{i=1}^{\ell} m_i(m_i-1) -\frac{1}{2} 4E_2(m_1,\dotsc, m_{\ell})    \right)\\[2mm]\nonumber
			&= (-2)^{s(D_0)}\Bigg( -\sum_{i=1}^{\ell} m_i - \underbrace{ \left( \sum_{i=1}^{\ell} m_i^2 + 2E_2(m_1,\dotsc, m_{\ell}) \right) }_{=\left(\sum_{i=1}^{\ell} m_i\right)^2 = m^2 } \Bigg)\\[1mm]
			&= -m( m + 1)\cdot(-2)^{s(D_0)}.
		}
		
	\end{proof}
	\end{theo}

	Inspired by the above theorem we can now consider iterations $\bar{\cD}(\bar{\cD}(D_0)), \dotsc, \bar{\cD}^{m}(D_0)$, i.e. sums over sets of chord diagrams, which result from adding multiple chords in all possible ways.
	\begin{cor}
		Let $D_0 \in \cD_{N-m}^n$ with $1\leq m\leq N = \sum n_i$ and $\bar{\cD}(D_0)$ as in theorem \ref{theo_sumTrace} above. Then, for $1\leq k \leq m$
		\al{
			\sum_{ D\in \bar{\cD}^k(D_0) } (-2)^{s(D)} = (-1)^kk!\binom{m}{k}\binom{m+1}{k} (-2)^{s(D_0)}.
		}
		In particular, one finds the sum over all completions of $D_0$ for $k=m$
		\al{
			\sum_{ D\in \bar{\cD}^m(D_0) } (-2)^{s(D)} = (-1)^m(m+1)! (-2)^{s(D_0)}
		}
		and the sum over all diagrams of a given order $2N$ for $k=m=N$
		\al{
			\sum_{ D\in \cD^n_N} (-2)^{s(D)} = 4(-1)^N(N+1)!
		}
		as the completions of the empty chord diagram on $2N$ vertices.
	\begin{proof}
		For $k=1$ the statement is just theorem \ref{theo_sumTrace} since $\binom{m}{1}\binom{m+1}{1} = m(m+1)$. For $k>1$ we make use of theorem \ref{theo_sumTrace} iteratively, collect the factors and divide by $k!$ to account for the different permutations of chord additions that result in the same diagrams:
		\al{\nonumber
			\sum_{ D\in \bar{\cD}^k(D_0) } (-2)^{s(D)}
				&=\frac{1}{k!} \sum_{ D_1\in \bar{\cD}(D_0) }\nquad\dotsm
										\sum_{ D_k\in \bar{\cD}(D_{k-1}) } \nquad(-2)^{s(D_k)}\\[2mm]\nonumber
				&=\frac{(-1)}{k!}(m-k+1)(m-k+2) \sum_{ D_1\in \bar{\cD}(D_0) }\nquad\dotsm
										\sum_{ D_{k-1}\in \bar{\cD}(D_{k-2}) } \nquad(-2)^{s(D_{k-1})}\\[2mm]\nonumber
				&\quad \vdots\\\nonumber
				&=\frac{(-1)^k}{k!} \left(\prod_{i=1}^{k} (m-k+i)(m-k+i+1) \right) (-2)^{s(D_0)}\\[2mm]\nonumber
				&=\frac{(-1)^k}{k!} \frac{m!}{(m-k)!}\frac{(m+1)!}{(m-k+1)!} (-2)^{s(D_0)}\\[2mm]
				&=(-1)^kk!\binom{m}{k}\binom{m+1}{k} (-2)^{s(D_0)}
		}
	\end{proof}
	\end{cor}
	
\subsection{Summation with polynomials}
	\begin{conj}\label{conj_sumTrace}
		Let $D_0\in \cD^n_0$ with $n\in \NN^{\ell}$ and $N=\sum n_i$. Then
		\al{
			\sum_{D\in \cD^n_N} (-2)^{s(D)} \nquad\prod_{(u,v)\in E_D^0} \cp{u}{v} = (-2)^{\ell}\Big( Z_{12}(D_0) + Z_{21}(D_0) \Big).
		}
		where
		\al{
			Z_{ij}(D_0) \defeq \sum_{\cE \in \cP(E_{D_0}^i)} (-\kp)^{N-|\cE|} (|\cE|+1)!\ Y(\cE, E_{D_0}^j).
		}		
		$\cP(E_{D_0}^i)$ here denotes the set of partitions $\cE=(E_1,\dotsc, E_{|\cE|})$ of $i$-coloured base edges and $Y(\cE, E_{D_0}^j)$ is a sum of products of $|\cE|$ Dodgson polynomials $\Psi_{\Gamma, K}^{I,J}$ with $|I| = |E_k| = |J|$ and $K=\emptyset$, associated to the underlying graph $\Gamma$.
	\end{conj}

	A number of remarks regarding this conjecture and potential future work are in order:
	\begin{itemize}
		\item The Dodgson polynomials were introduced by Brown \cite{Brown_2010_Periods}. They enter this setting because, up to a sign ambiguity, $\cp{i}{j} = \pm\Psi_{\Gamma, \emptyset}^{\{i\},\{j\}}$. A potential proof of the conjecture should then rely on the Dodgson identity
			\al{
				\Psi_{\Gamma, \emptyset}^{\{i_1\},\{j_1\}}\Psi_{\Gamma, \emptyset}^{\{i_2\},\{j_2\}}
				-\Psi_{\Gamma, \emptyset}^{\{i_1\},\{j_2\}}\Psi_{\Gamma, \emptyset}^{\{i_2\},\{j_1\}}
				= \kp \Psi_{\Gamma, \emptyset}^{\{i_1,i_2\},\{j_1,j_2\}}
			}
			and its higher order generalisations.
		\item The precise definition of $Y(\cE, E_{D_0}^j)$ requires rather extensive exposition beyond the scope of this article and shall be given in future work, together with a rigorous proof. Such a proof, using the Dodgson identity, seems in reach in principle but is an enormous combinatorial mess that needs to be worked out in detail elsewhere.
		\item The conjecture has been checked computationally for all possible configurations $n = (n_1,\dotsc, n_{\ell})$ up to and including $\sum n_i=7$, in which it is a sum over $(2\cdot7-1)!! = 135135$ chord diagrams.
		\item The chord sum corresponding to $k=n$ in the integrand \refeq{eq_num} can be simplified by this conjecture. Moreover, due to transversality the $k=n-1$ term should yield the exact same result when integrating and it should be possible to prove this directly on the level of the integrand (by showing that they yield integrands that are equal up to exact forms), observing that the various $\cp{i}{j}$ and $X_{\Gamma}^{i,\mu_i}$ are all first or second derivatives of $\Phi_{\Gamma}$ \cite{Golz_2017_CyclePol}.
		\item The remaining terms with $k<n-1$ are all convergent and vanish in renormalisation, at least at the superficial level. For graphs with subdivergences there is a rather complicated interplay between convergent and divergent parts of sub- and cographs that needs to be studied in detail.
		\item The $(-\kp)^{N-|\cE|}$ cancels some powers of the Kirchhoff polynomial in the denominator of the integrand, massively reducing its size (in terms of Schwinger parameter monomials) as well es computational complexity. While the overall transcendental degree of the integrand remains the same, the number of terms with the highest power of $\kp$ in the denominator is reduced to the two expressions
		\al{
			\prod_{(u,v)\in E_{D_0}^1} \cp{u}{v} + \prod_{(u,v)\in E_{D_0}^2} \cp{u}{v}
		}
		corresponding to the two partitions that separate $E_{D_0}^1$ and $E_{D_0}^2$ into $N$ parts containing only one edge each.
	\end{itemize}
	
\newpage
\section*{Conclusion}
	The process of contracting traces of Dirac matrices was abstracted to a purely combinatorial level. Using methods that revealed possible interdisciplinary connections to theoretical computer science, we found a formula that replaces the contraction entirely and expresses the end result in terms of the structure of chord diagrams associated to traces of Dirac matrices. This allowed us to rewrite the complicated numerator structure of parametric Feynman integrands in quantum electrodynamics as a sum over chord diagrams, with all contractions fully executed. An even further simplified expression was conjectured, based on extensive example computations and the properties of the graph polynomials appearing in the integrand. Due to an abundance of cancellations and the elimination of the Dirac matrix structures this conjectured expression massively reduces the overall size of the integrand, making it accessible to automated integration by a computer, potentially to higher loop numbers than before. Additionally, the simplified structure of the integrand opens it up to algebro-geometric and number theoretic studies, e.g. regarding the appearance and cancellation of transcendentals in QED amplitudes \cite{BroadhurstDelbourgoKreimer_1996_Unknotting, GKLS_1991_QED, Rosner_1966_QED}.\\

	The method applies to any QED Feynman graph. Moreover, it should be possible -- albeit combinatorially more complex -- to generalise the work presented here to QCD or general gauge theories. Consider the general Schwinger parametric integrand for gauge theories as derived in \cite{KreimerSarsSuijlekom_2013_QuantGauge}. It reduces Feynman graphs with 4-valent vertices to sums of 3-regular graphs, making it analogous to the QED case discussed here, and uses the so-called ``Corolla differential'' to express the numerator structure. This differential is a generalisation of the derivatives discussed in \cite{Golz_2017_CyclePol} and computing it in terms of Dodgson polynomials appears to be mostly a matter of sorting through large numbers of derivatives and applying identities already proved in \cite{Golz_2017_CyclePol}. The reduction to 3-regular graphs suggests that general gauge theories can be treated similarly to QED Feynman graphs, although it remains to be seen how exactly the contraction formalism has to be modified to incorporate some of the more complicated objects arising from non-abelian gauge theories.

	\bibliographystyle{plainurl}
	\bibliography{/Users/marcelgolz/Documents/University/LaTeX/Bibliography/bibliography}

\end{document}